\documentclass[11pt]{article}
\usepackage{amssymb,amsmath,amsfonts}
\usepackage{graphicx}
\usepackage{graphics}
\usepackage{eepic,epsfig}

1.60cm \makeatletter \@addtoreset{equation}{section}

\makeatother

\begin{document}

\title{Finite temperature bosonic charge and current densities in \\ compactified
cosmic string spacetime}
\author{A. Mohammadi$^{1}$\thanks{%
E-mail: a.mohammadi@fisica.ufpb.br}  \, and \, E. R. Bezerra de Mello $^{1}$\thanks{%
E-mail: emello@fisica.ufpb.br} \\
\\
\textit{$^{1}$Departamento de F\'{\i}sica, Universidade Federal da Para\'{\i}%
ba}\\
\textit{58.059-970, Caixa Postal 5.008, Jo\~{a}o Pessoa, PB, Brazil}\vspace{%
0.3cm}\\
}
\maketitle

\begin{abstract}
In this paper we study the expectation values of the induced charge and
current densities for a massive bosonic field with nonzero chemical potential in the geometry of a higher dimensional compactified cosmic string with magnetic fluxes, along the string core and also enclosed by the compactified direction, in thermal equilibrium at finite temperature $T$. These densities are calculated by decomposing them into the vacuum expectation values and finite temperature contributions coming from the particles and antiparticles. The
only nonzero components correspond to the charge, azimuthal and axial current densities. By using the Abel-Plana formula, we decompose the components of the densities into the part induced by the cosmic string and the one by the compactification.
The charge density is an odd function of the chemical potential and even periodic function of the magnetic flux with a period equal to the quantum flux. Moreover, the azimuthal (axial) current density is an even function of the chemical potential and an odd (even) periodic function of the magnetic flux with the same period. In this paper our main concern is the thermal effect on the charge and current densities, including some limiting cases, the low and high temperature approximations. We show that in all cases the temperature enhances the induced densities.

\end{abstract}

\bigskip


\bigskip

\section{Introduction}
The existence of a magnetic flux tube penetrating a type II superconductor, named ``vortex'', was first demonstrated by Abrikosov \cite{Abrikosov}, by using the Ginzburg-Landau theory of superconductivity. 
A few years later, Nielsen and Olesen 
proposed a relativistic theoretical model consisting of Higgs and gauge fields, that possesses static vortex solution. In this model the Higgs field plays the role of the superconductivity order parameter\cite{Nielsen}. The influence of the vortex from the Nielsen-Olesen model on the geometry of the spacetime was investigated by Garfinkle \cite{Garfinkle}. The author showed that the spacetime around the vortex is asymptotically a
Minkowski one minus a wedge. The core of the vortex has a nonzero
thickness and a magnetic flux through it. A few years later, Linet \cite{Linet} showed that, under specific condition, the structure of the respective spacetime corresponds to a conical one, with the conicity parameter being expressed in terms of the energy per
unit length of the vortex. Meanwhile, Kibble has worked on mechanisms of symmetry breaking, phase transitions and the topological defects and introduced the phenomenon into the modern cosmology\cite{Kibble}.

Cosmic strings are lines of trapped energy density, analogous to defects such as vortex lines
in superconductors and superfluids.  An enormous number of string type solutions have been found
in many different field theoretic models, including electroweak strings in Weinberg-Salam
model and strings in grand unified theories \cite{Jeannerot}. 
Accepting the validity of grand unified theories, one can conclude that in the earliest
stages, the universe was hotter and in a
more symmetric state. Thus, in its expansion
process the
universe cooled down and underwent a series of phase transitions accompanied
by spontaneous breakdown of symmetries. As a result, topological
defects could then be left as remnants.

Though the recent observational data from
Planck and WMAP have ruled out cosmic strings as the primary source for primordial density
perturbations, they have attracted a renewed interest in the context of string cosmology.
Sarangi and Tye predicted the production of cosmic superstrings during the last stages of brane inflation \cite{Tye}. In the context of brane inflation various formation mechanism has been proposed \cite{Polchinski}. The qualitative properties
of such cosmic superstrings in the late time universe should be similar to those of field theoretic
strings, except for the fact that the intercommuting probability is relatively low for cosmic superstrings. 

Cosmic strings can curve spacetime and be of cosmological
and astronomical significance in a large number of phenomena, such as
producing cosmic microwave background anisotropies, non-Gaussianity and B-mode polarization, sourcing gravitational waves, generation of high energy cosmic rays and gravitationally lensing astrophysical objects \cite{Hindmarsh2}. 
The dimensionless parameter that characterizes the strength of
gravitational interactions of strings with matter is its ``tension'' in Planck units, $G\mu_0$, where $G$
is Newton's constant and $\mu_0$
is the mass per unit length, proportional to the square of the symmetry breaking scale.

Many authors have considered quantum fields on spacetimes containing conical singularities. These kind of manifolds 
are relevant for studying fields in the presence of real conical singularities, namely existing in the 
Lorentzian section of the manifold as in the case of idealized cosmic strings. Moreover, the presence of 
the string allows effects such as  particle-antiparticle pair production by a single photon and bremsstrahlung 
radiation from charged particles which are not possible in empty Minkowski space due to the conservation
of linear momentum \cite{Skarzhinsky}. 

Cosmic strings also polarize
the vacuum around it, in a way similar to the distortion of the vacuum leading to
the Casimir effect between two conducting planes forming a wedge.

Another type of topological effects that is considered in the literature is induced
by the compactification of the spatial dimension. The compact spatial dimensions are an inherent feature of most
high energy theories of fundamental physics, including supergravity and
superstring theories. An interesting application of the field theoretical
models with compact dimensions recently appeared in nanophysics. The
long wavelength description of the electronic states in graphene can be
formulated in terms of the Dirac like theory in three dimensional spacetime
with the Fermi velocity playing the role of the speed of light \cite{Cast09}. In addition to fermionic field, the scalar and gauge fields originated from the elastic properties and describing disorder phenomena, like the distortion of graphene lattice and structure defects should be taking into consideration \cite{Jackiw,Oliveira}. In graphene made structures, like cylindrical and toroidal carbon
nanotubes, the background geometry for the corresponding field theory
contains one or two compact dimensions. In quantum field theory, the
periodicity conditions imposed on the field operator along compact
dimensions modify the spectrum for the normal modes and as a result
the vacuum expectation values of physical observables are changed. 
In particular, many authors have investigated the vacuum energy and stresses
induced by the presence of compact dimensions (for reviews see Refs. \cite{Most97}, \cite{Duff86}). This effect, known as topological Casimir effect,
is a physical example of the connection between global properties of the
spacetime and quantum phenomena. 

The analysis of the vacuum polarization by a magnetic flux tube at finite temperature in the cosmic string spacetime has been developed in \cite{Spinelly}. In \cite{JPA}, we have analyzed the fermionic charge and current densities at finite temperature in a (3+1)-cosmic string spacetime considering the presence of a magnetic flux running along its axis. Here we have decided to continue in the same line of investigation considering a charged scalar field in a more general situation, in arbitrary dimensions and compactification along $z$ axis, for the bosonic field. 

The plan of the paper is as follows: In the next section, we introduce the background geometry and describe the thermal Hadamard function for the massive charged bosonic quantum field in a higher dimensional cosmic string spacetime presenting a magnetic flux along its axis. Moreover, we consider that the $z$-axis along the string is compactified to a circle and carries an extra magnetic flux. Also, in section \ref{Charge density}, we calculate the thermal average of the charge density in this spacetime. We investigate various asymptotic limits in detail, including the low and high temperature limits. In this paper we mainly explore the thermal effects on the average of the current density. The only nonzero components of the thermal average of the current densities correspond to the azimuthal and axial ones. These quantities are investigated in section \ref{Azimuthal current} and \ref{Axial current}. We show that the
expectation value of the current densities can be decomposed into the vacuum expectation value and the contributions from particles and antiparticles. The behavior of the current densities in the asymptotic regions of the parameters are discussed in detail. Finally, in section \ref{conclusion} we give a brief conclusion about the results.

\section{Formalism}

\label{sec1}
In the general relativity framework, the background geometry of the $(D + 1)$ dimensional flat spacetime corresponding to a generalized cosmic string lying along the z-axis can be described by the line element
\begin{equation}
ds^{2}=dt^{2}-dr^{2}-r^{2}d\phi ^{2}-dz{}^{2}-d\bold{z}_{||}{}^{2}\ ,  \label{ds21}
\end{equation}%
where $r\geqslant 0$, $0\leqslant
\phi \leqslant \phi _{0}=2\pi /q$, $-\infty <t<+\infty $, $d\bold{z}_{||}{}^{2}=\sum_{i=4}^{D} (d x ^i)^2$ and $-\infty <x ^i <+\infty $.  The parameter  $q$
codifies the planar angle deficit where in the case of $D=3$ is related to the linear mass density
of the string $\mu_{0}$ by $q^{-1}=1-4\mu _{0}$.
In the presence of a gauge field, $A_{\mu }$, the field equation governing the quantum dynamics of the charged massive scalar field is
\begin{equation}
\left({\mathcal{D}}^2+m^2\right)\varphi(x)=0,  \label{K-G}
\end{equation}%
where the differential operator in the field equation reads
\begin{align}
{\mathcal{D}}^2=\frac{1}{\sqrt{|g|}}{\mathcal{D}}_{\mu}\left(\sqrt{|g|}g^{\mu \nu} {\mathcal{D}}_{\nu
}\right), \ {\mathcal{D}}_{\mu
}=\partial _{\mu }+ieA_{\mu }\ .  \label{1}
\end{align}%
We assume that the direction along the $z$-axis is compactified to a circle with length $L$, $%
0\leqslant z\leqslant L$ and the field obeys quasiperiodic boundary condition
\begin{equation}
\varphi (t,r,\phi ,z+L,\bold{z}_{||})=e^{2\pi i\eta }\varphi (t,r,\phi ,z,\bold{z}_{||})\ ,  \label{Period}
\end{equation}%
 with the constant phase $\eta $ in the region $[0,1]$.  We shall consider a constant vector potential as
\begin{equation}
A_{\mu }=(0,0,A_{\phi },A_{z},0,...,0)\ .  \label{Amu}
\end{equation}%
The azimuthal component $A_{\phi }$ is related to an infinitesimal magnetic flux,
$\Phi _{\phi }$, running along the string as $A_{\phi }=-q\Phi _{\phi
}/(2\pi )$ and the axial component $A_{z}$ is related to the magnetic flux $\Phi _{z}$ enclosed by the $z$-axis given by $A_{z}=-\Phi _{z}/L$. It is easy to show that in the presence of nonzero constant vector potential components $\bold{A}_{\bold{z_{||}}}$, the physical parameters do not change and the results will remain intact. Therefore, without loss of generality we shall consider $\bold{A}_{\bold{z_{||}}}=0$. As it is shown in \cite{Eduardo}, the general normalized solution in this system takes the form
\begin{equation}
\varphi_\sigma(x)=\left(\frac{q\lambda}{2(2\pi)^{D-2} E_l L}\right)^{1/2} J_{q|n+\alpha|} (\lambda r) e^{-iE_l t +iqn\phi+ik_l z+i\bold k_{||}. \bold z_{||}},
\end{equation}
where $\sigma$ represents the set of quantum numbers $(n, \lambda, k_l, \bold k_{||})$, $J_{\nu }(x)$ the Bessel function, $n=0,\pm1,\pm2,\ldots$, $\lambda \geq 0$, $k_l=2\pi(l+\eta)/L$ with $l=0,\pm 1,\pm 2\ ,\ldots$, $-\infty<k^j_{||}<\infty$ with $j=4,...,D$ and
\begin{equation}
\;\alpha =eA_{\phi }/q=-\Phi
_{\phi }/\Phi _{0},  \label{betaj}
\end{equation}%
$\Phi _{0}=2\pi /e$ being the quantum flux.
The energy is expressed in terms of $\lambda $ and $l$ by the relation
\begin{equation}
E_l=\sqrt{\lambda ^{2}+\tilde{k}_{l}^{2}+\bold{k}_{||}^2+m^{2}}\ ,\;\tilde{k}_{l}=2\pi (l+
\tilde{\eta})/L, \ \ l=0,\pm 1,\pm 2\ ,\ldots \ ,  \label{E+}
\end{equation}
where
\begin{equation}
\tilde{\eta}=\eta +eA_{z}L/(2\pi )=\eta -\Phi _{z}/\Phi _{0}\ .
\label{bett}
\end{equation}
The momentum along the string axis is discrete due to the compactification along its axis.

In order to obtain the expectation values of the charge and current densities for above bosonic field at finite temperature T, we shall use the thermal Hadamard function
\begin{equation}
G^{(1)}(x,x') =Tr\left[\hat{\rho}\left(\varphi^{*}(x')\varphi(x)+\varphi(x) \varphi^{*}(x') \right)\right] \ ,  \label{C}
\end{equation}%
where $\hat{\rho}$ is the density matrix
\begin{equation}
\hat{\rho}=Z^{-1}e^{-\beta (\hat{H}-\mu ^{\prime }\hat{Q})},\;\beta =1/T.
\label{rho}
\end{equation}%
In this equation $\hat{H}$ is the Hamiltonian operator, $\hat{Q}$ a conserved charge, $\mu ^{\prime }$ the corresponding chemical potential, and 
\begin{equation}
Z=\mathrm{tr}[e^{-\beta (\hat{H}-\mu ^{\prime }\hat{Q})}] \ , \label{partition}
\end{equation}%
 is the grand canonical partition function. 
Expanding the field operator over a complete set of solutions as
\begin{equation}
\varphi(x)=\sum_{\sigma }[\hat{a}_{\sigma }\varphi _{\sigma }^{(+)}(x)+\hat{b}_{\sigma
}^{+}\varphi _{\sigma }^{(-)}(x)] \ ,  \label{psiexp}
\end{equation}%
and using the following relations
\begin{eqnarray}
\mathrm{tr}[\hat{\rho }\hat{a}_{\sigma }^{+}\hat{a}_{\sigma ^{\prime }}] &=&%
\frac{\delta _{\sigma \sigma ^{\prime }}}{e^{\beta (E_l-\tilde{\mu} )}-1},  \notag \\
\mathrm{tr}[\hat{\rho }\hat{b}_{\sigma }^{+}\hat{b}_{\sigma ^{\prime }}] &=&%
\frac{\delta _{\sigma \sigma ^{\prime }}}{e^{\beta (E_l+\tilde{\mu} )}-1} ,  \label{traa}
\end{eqnarray}%
where $\tilde{\mu}=e \mu'$ and $\{\varphi _{\sigma
}^{(+)},\varphi _{\sigma }^{(-)}\}$ a complete set of normalized positive and negative energy solutions  of Eq. (\ref{K-G}), one can obtain the expectation value in Eq. (\ref{C}). 

The bosonic current density operator is given by
\begin{equation}
j_{\nu } =i e\left[\varphi^{*}(x){\mathcal{D}}_{\nu
} \varphi(x)-\varphi(x) \left ({\mathcal{D}}_{\nu}\varphi(x)\right)^{*}\right]\ .  \label{C2}
\end{equation}%
The thermal average of the above current density can be obtained in terms of thermal Hadamard function as 
\begin{equation}
\left\langle j_{\nu }\right\rangle =i e \ \mathrm{Lim}_{x' \rightarrow x} \left[ (\partial_\nu-\partial'_\nu+2i e A_\nu) G^{(1)}(x,x')\right]\ .  \label{Cur}
\end{equation}%

Substituting the bosonic field operator in Eq. (\ref{C}) by the expansion (\ref{psiexp}), and using the relations in Eq. (\ref{traa}), we can separate the zero temperature Hadamard function and the one related to particles and antiparticles as shown below
\begin{equation}
G^{(1)}(x,x')=G_0^{(1)}(x,x')+G^{(1)}_T(x,x'), \label{1}
\end{equation}
where $G_0^{(1)}(x,x')$ is the zero temperature Hadamard function and given by
\begin{eqnarray}
&&G_0^{(1)}(x,x')=\sum_{\sigma }\sum_{\chi=+,-}\varphi_{\sigma}^{(\chi)}(x) \varphi_{\sigma}^{(\chi) *}(x')=\frac{q}{2 (2 \pi)^{D-2} L} \sum_{\sigma }\ \lambda J_{q|n+\alpha|}(\lambda r)J_{q|n+\alpha|}(\lambda r') \notag \\
&&\times \frac{e^{iqn\Delta \phi+i\tilde{k}_l \Delta z+i \bold k_{||}.\Delta \bold z_{||}}}{E_l}\left\{e^{-i E_l \Delta t}+e^{i E_l \Delta t}\right\}, \label{1}
\end{eqnarray}
and the one related to the particles and antiparticles which is given as follows
\begin{eqnarray}
&&G^{(1)}_T(x,x')=2 \sum_{\sigma }\sum_{\chi=+,-}\frac{\varphi_{\sigma}^{(\chi)}(x) \varphi_{\sigma}^{(\chi) *}(x')}{e^{\beta (E_l-\chi
\tilde{\mu})}-1}=\frac{q}{ (2 \pi)^{D-2} L} \sum_{\sigma }\ \lambda J_{q|n+\alpha|}(\lambda r)J_{q|n+\alpha|}(\lambda r') \notag \\
&&\times \frac{e^{iqn\Delta \phi+i\tilde{k}_l \Delta z+i \bold k_{||}.\Delta \bold z_{||}}}{E_l}\left\{\frac{e^{-i E_l \Delta t}}{e^{\beta(E_l-\tilde{\mu})}-1}+\frac{e^{i E_l \Delta t}}{e^{\beta(E_l+\tilde{\mu})}-1}\right\}. \label{1}
\end{eqnarray}
Due to this separation for thermal Hadamard function, the same will happen for the vacuum expectation value of the current density 
\begin{equation}
\left\langle j_{\nu }\right\rangle =\left\langle j_{\nu }\right\rangle
_{0}+\left\langle j_{\nu }\right\rangle
_{T} \, \label{Curr}
\end{equation}%
where $\left\langle j_{\nu }\right\rangle_{T}=\sum _{\chi=+,-}\left\langle j_{\nu }\right\rangle
_{\chi}$ is sum of contributions of particles and antiparticles in the current density, originating from the nonzero temperature.
The bosonic charge and current densities at zero temperature, have been
investigated in \cite{Eduardo} for zero chemical potential. Therefore, we are mainly concerned with
the contributions from particles and antiparticles provided by the second term in the right-hand side of (\ref{Curr}). In the limit $T\rightarrow 0$ we expect this term vanishes and only the first term, the zero temperature contribution, survives.

\section{Charge density}
\label{Charge density}

First we consider the charge density corresponding to $\nu=0$ component of Eq. (\ref{Cur}). Substituting the thermal Hadamard function in this equation and concerning just the contributions from particles and antiparticles, we have
\begin{equation}
\left\langle j_{0}\right\rangle _{T}=\frac{eq}{(2\pi)^{D-2} L}\sum_{\sigma }\ \lambda \ J_{q|n+\alpha|}^{2}(\lambda r) \ \left[\frac{1}{e^{\beta(E_l-\tilde{\mu})}-1}-\frac{1}{e^{\beta(E_l+\tilde{\mu})}-1}\right],
  \label{Chargeden}
\end{equation}
with the notation
\begin{equation}
\sum_{\sigma }=\int d\bold{k}_{||}\int_{0}^{\infty }d\lambda \ \sum_{l=-\infty }^{+\infty
}\sum_{n=-\infty}^{+\infty}\ .  \label{Sumsig}
\end{equation}
It has been shown in \cite{Eduardo} that the vacuum expectation value of the charge density vanishes, $\left\langle j_{0}\right\rangle _{0}=0$. As can be seen, the charge density is an odd function of $\tilde{\mu}$. When the chemical potential $\tilde{\mu}$ is zero the contributions from the particles and antiparticles cancel each other and the total charge density vanishes. The presence of the nonzero chemical potential imbalances the particle-antiparticle contributions and creates nonzero charge density. In Eq. (\ref{Chargeden}), the charge density is an even periodic function of the parameter $\alpha$ with the period equal to 1 which means that it is a periodic function of the
magnetic flux with a period equal to the quantum flux. Presenting the parameter $\alpha$ as
\begin{equation}
\alpha=n_0+\alpha_0 \ ,  \label{Aharonov}
\end{equation}
where $n_0$ is the integer part and $\alpha_0$ the fractional part of $\alpha$, choosing $|\alpha_0|\leqslant1/2$,  it will be shown shortly that the bosonic charge and current densities are only dependent on the fractional part $\alpha_0$ which resembles Aharonov-Bohm effect.

Notice that, the bosonic chemical potential, in contrast to the fermionic one which can have any value in principle, is restricted by $|\tilde{\mu}|\leqslant \epsilon_0$, $\epsilon_0$ being the minimum of energy. Therefore, $E_l\pm \tilde{\mu}$ is always positive and we can simply use the following series expansion
\begin{equation}
(e^{y}-1)^{-1}=\sum_{j=1}^{\infty }e^{-jy} \ ,  \label{Expansion}
\end{equation}
in our analysis.
Using the above expansion in (\ref{Chargeden}), we have
\begin{align}
\left\langle j_{0}\right\rangle _{T}=\frac{2eq}{(2\pi)^{D-2} L}\int d\bold{k}_{||}\int^\infty_0 d\lambda \ \sum_{l=-\infty }^{+\infty
}\sum_{n=-\infty}^{+\infty}\ \lambda \ J_{q|n+\alpha|}^{2}(\lambda r) \sum_{j=1}^\infty  e^{-j \beta E_l} \sinh(j \beta \tilde{\mu}) \ .
  \label{3}
\end{align}%
In order to develop the summation over $l$, we apply the
Abel-Plana summation formula in the form \cite{SahaRev,Beze08}
\begin{eqnarray}
&&\sum_{l=-\infty }^{\infty }g(l+\tilde{\eta})f(|l+\tilde{\eta}%
|)=\int_{0}^{\infty }du\,\left[ g(u)+g(-u)\right] f(u)  \notag \\
&&\qquad +i\int_{0}^{\infty }du\left[ f(iu)-f(-iu)\right] \sum_{\lambda =\pm
1}\frac{g(i\lambda u)}{e^{2\pi (u+i\lambda \tilde{\eta})}-1}\ ,
\label{sumform}
\end{eqnarray}%
which helps to separate the part in the charge density induced by the compactification. Taking
$g(u)=1$ and
\begin{equation}
f(u)=e^{-j \beta \sqrt{\lambda ^{2}+(2\pi u/L)^{2}+\bold{k}_{||}^2+m^{2}}},  \label{fu}
\end{equation}%
by using (\ref{sumform}), we can write the charge density as
\begin{equation}
\left\langle j_{0}\right\rangle _{T} =\left\langle j_{0}\right\rangle _{Ts}+\left\langle j_{0}\right\rangle _{Tc}\ ,
\end{equation}%
with
\begin{align}
\left\langle j_{0}\right\rangle _{Ts}=\frac{eq}{(2\pi)^{D-2} \pi}\int d\bold{k}_{||}\int^\infty_0 d\lambda \ \int_{-\infty }^{+\infty} \ d k \sum_{n=-\infty}^{+\infty}\ \lambda \ J_{q|n+\alpha|}^{2}(\lambda r) \sum_{j=1}^\infty  e^{-j \beta E_l} \sinh(j \beta \tilde{\mu}) \ ,
  \label{3}
\end{align}%
being the contribution induced by the string and
\begin{align}
\left\langle j_{0}\right\rangle _{Tc}&=\frac{2eq}{(2\pi)^{D-2} \pi}\int d\bold{k}_{||}\int^\infty_0 d\lambda \   \int^\infty_{\sqrt{\lambda ^{2}+\bold{k}_{||}^2+m^{2}}} dk \sum_{n=-\infty}^{+\infty}\ \lambda \  J_{q|n+\alpha|}^{2}(\lambda r) \nonumber\\
&\times \sum_{j=1}^\infty  \sinh(j \beta \tilde{\mu}) \sin\left({j \beta \sqrt{k^{2}-\bold{k}_{||}^2-\lambda ^{2}-m^{2}} }\right) \left[\frac{1}{e^{L k+2\pi i \tilde{\eta}}-1}+\frac{1}{e^{L k-2\pi i \tilde{\eta}}-1}\right] \ ,
  \label{chdencom}
\end{align}%
due to the compactification. As we shall see, this contribution goes to zero in the limit $L\rightarrow \infty$.

Let us first calculate the string part. In order to make it workable, we use the integral representation below
\begin{equation}
e^{-j\beta\sqrt{\lambda ^{2}+k^2+\bold{k}_{||}^2+m^{2}}}=\frac{j \beta}{\sqrt{\pi }}\int_{0}^{\infty
}ds\,s^{-2}e^{-(\lambda ^{2}+k^2+\bold{k}_{||}^2+m^{2})s^{2}-j^2 \beta^2/4s^{2}}. \label{IntRep2}
\end{equation}%
So, we can calculate the integral over $\lambda$ by using \cite{Grad}
\begin{equation}
\int_{0}^{\infty }d\lambda \ \lambda \ e^{-\eta \lambda ^{2}}J_{\nu}^{2}(\lambda r) =%
\frac{e^{-r^{2}/(2\eta )}}{2\eta } I_{\nu}(r^{2}/(2\eta
)) \ .  \label{Int-reg}
\end{equation}%
The integral over $\bold{k}_{||}$ is easily done as 
\begin{equation}
\int d\bold{k}_{||} e^{-s^2 \bold{k}_{||}^2}=(\pi/s^2)^{(D-3)/2} \ .  \label{extrad}
\end{equation}%
Therefore, we obtain the following result for the string contribution of the charge density
\begin{align}
\left\langle j_{0}\right\rangle _{Ts}&=\frac{4eq \beta}{(4\pi)^{(D+1)/2}} \ \int^\infty_0 ds\  s^{-D-2} F(q,\alpha_0,r^2/2 s^2) \nonumber\\
&\times \sum_{j=1}^\infty j   \sinh(j \beta \tilde{\mu}) \ e^{-m^2 s^2-(j^2 \beta^2/4+r^2/2)s^{-2}} \ ,
  \label{6}
\end{align}%
where we have introduced \cite{Eduardo}
 \begin{align}%
 F(q,\alpha_0,z)& = \sum_{n=- \infty}^{+\infty}\ I_{q|n+\alpha|}(z) \nonumber\\
&=\frac{1}{q}\left[ e^{z}-\frac{q}{\pi }%
\int_{0}^{\infty }dy \ \frac{e^{-z\cosh {(y)}}h(q,\alpha
_{0},y)}{\cosh (qy)-\cos (q\pi )}
+2\sideset{}{'}\sum_{k=1}^{[q/2]}\cos \left( 2\pi k\alpha _{0}\right)
e^{z\cos (2\pi k/q)}\right] \ ,  \label{roReg1}
 \end{align}%
 with the notation
\begin{equation}
h(q,\alpha_{0},y)=\sin \left[ \left( 1- |\alpha_{0}| \right) q\pi %
\right] \cosh \left( |\alpha_{0}|  qy\right)+\sin \left( |\alpha_{0}| q\pi %
\right) \cosh \left[ \left( 1- |\alpha_{0}|\right) qy\right] .  \label{h}
\end{equation}%
Note that $[q/2]$ stands for the integer part of $q/2$ and the prime on the summation in (\ref{roReg1}) and hereafter means that the term $k=q/2$ should be taken with the coefficient $1/2$ when $q$ is an even number. For $q<2$ the summation on $k$ does not contribute.

Substituting the expression found for $ F(q,\alpha_0,r^2/2s^2)$ into (\ref{6}) and integrating over parameter $s$, using the following relation
\begin{equation}
\int_0^{\infty}ds \ s^{-d} e^{-a s^2-bs^{-2}}=\left(\frac{a}{b}\right)^{(d-1)/4} K_{(d-1)/2}(2\sqrt{a b}) \ , \label{int s}
\end{equation}
leads to the final result for the thermal average of the charge density as
\begin{align}
\left\langle j^{0}\right\rangle _{Ts}&=\frac{4 e \beta m^{D+1}}{(2\pi)^{(D+1)/2} } \sum_{j=1}^\infty j \sinh(j \beta \tilde{\mu}) \nonumber\\
& \times \left \{ f_{(D+1)/2}(j m \beta)+2\sideset{}{'}\sum_{k=1}^{[q/2]}\cos \left( 2\pi k\alpha _{0}\right)\right. f_{(D+1)/2}\left(2 m c_{j,k}(\beta,q)\right)\nonumber\\
&-\frac{q}{\pi}\int_{0}^{\infty }dy \ \frac{ h(q,\alpha_{0},y)}{\cosh (qy)-\cos (q\pi )} \left. f_{(D+1)/2}\left(2 m s_j(\beta,y)\right) \phantom{\sum_{k=1}^{[q/2]}}\hspace{-0.7cm} \right \} \ .
  \label{densityfin}
\end{align}%
In the above relation we have introduced the following notations
\begin{align}
f_\nu(x)=\frac{K_\nu(x)}{x^\nu} \ , c_{j,k}(\beta,q)=\sqrt{j^2 \beta^2/4+r^2 \sin^2(\pi k/q)} \ , s_j(\beta,y) = \sqrt{j^2 \beta^2/4+r^2 \cosh^2 (y/2)}\ , \label{fnu}
\end{align}
$K_\nu(x)$ being the MacDonald function. The first term in (\ref{densityfin}) corresponds to the Minkowskian part, in the absence of the conical defect and magnetic flux ($q=1$ and $\alpha_0=0$). 

Now, let us calculate the induced charge density originating from the compactification. By making a change of variable
  as $p=\sqrt{k^2-\bold{k}_{||}^2-\lambda^2-m^2}$ in (\ref{chdencom}), we obtain
 \begin{align}
\left\langle j_{0}\right\rangle _{Tc}&=\frac{4eq}{(2\pi)^{D-2} \pi }\sum_{l=1}^\infty \cos(2\pi l \tilde{\eta})\int \ d\bold{k}_{||}\int^\infty_0 d\lambda \  \lambda \int^\infty_0 dp \ p \ \frac{e^{-l L \sqrt{p^2+\bold{k}_{||}^2+\lambda^2+m^2}}}{\sqrt{p^2+\bold{k}_{||}^2+\lambda^2+m^2}}\nonumber\\
& \times \sum_{n=- \infty }^{+ \infty}\ J_{q |n+\alpha|}^{2}(\lambda r) \sum_{j=1}^\infty  \sinh(j \beta \tilde{\mu}) \sin\left({j \beta p }\right) \ .
  \label{7}
\end{align}
The following integral representation
\begin{equation}
\frac{e^{-lL\sqrt{\lambda ^{2}+p^{2}+\bold{k}_{||}^2+m^2}}}{\sqrt{\lambda ^{2}+p^{2}+\bold{k}_{||}^2+m^2}}=\frac{2}{%
\sqrt{\pi }}\int_{0}^{\infty }ds\,e^{-(\lambda
^{2}+p^{2}+\bold{k}_{||}^2+m^2)s^{2}-l^{2}L^{2}/4s^{2}},  \label{Rel4}
\end{equation}%
makes it possible to integrate over $p$, $\lambda$ and $\bold{k}_{||}$ analytically. The integration over $p$ is given below
\begin{align}
\int_0^\infty dp \ p \ e^{-p^2 s^2} \sin(j \beta p)=\frac{j \beta \sqrt{\pi}}{4 s^3} e^{-j^2 \beta^2/4 s^2} \ .
\end{align}
Integration over $\lambda$ and $\bold{k}_{||}$ will be the same as (\ref{Int-reg}) and (\ref{extrad}), respectively. Therefore, using the expression (\ref{roReg1}), the contribution of the compactification in the thermal charge density is
\begin{align}
\left\langle j^{0}\right\rangle _{Tc}&=\frac{8e \beta m^{D+1}}{(2\pi)^{(D+1)/2}}\sum_{l=1}^\infty \cos(2\pi l \tilde{\eta}) \sum_{j=1}^\infty j \sinh(j \beta \tilde{\mu}) \nonumber\\
& \times \left \{  f_{(D+1)/2}(m \rho_{j,l}(\beta))+2\sideset {}{'}\sum_{k=1}^{[q/2]}\cos \left( 2\pi k\alpha _{0}\right) \right.  f_{(D+1)/2}\left(2 m \sigma_{j,l,k}(\beta,q)\right) \nonumber\\
&-\frac{q}{\pi}\int_{0}^{\infty }dy \ \frac{h(q,\alpha_{0},y)}{\cosh (qy)-\cos (q\pi )}\left. f_{(D+1)/2}\left(2 m \delta_{j,l}(\beta,y)\right)\phantom{\sum_{k=1}^{[q/2]}}\hspace{-0.7cm}\right \} \ ,
  \label{comp-charge}
\end{align}%
where we have introduced the following notations
\begin{align}
\rho_{j,l}(\beta)=\sqrt{j^2 \beta^2+l^2 L^2} \ , \  \sigma_{j,l,k}(\beta,q)=\sqrt{j^2 \beta^2/4+l^2 L^2/4+r^2 \sin^2(\pi k/q)} \ , \nonumber \\  \delta_{j,l}(\beta,y)=\sqrt{j^2 \beta^2/4+l^2 L^2/4+r^2 \cosh^2 (y/2)} \ .
\label{rho-sigma-delta}
\end{align}

In the limit $L\rightarrow \infty$, using the asymptotic expansion for the MacDonald function for large arguments, we obtain
\begin{align}
\left\langle j^{0}\right\rangle _{Tc}&\approx \frac{4e \beta m^{D+1}}{(2\pi)^{D/2}}\sum_{l=1}^\infty \cos(2\pi l \tilde{\eta}) \frac{e^{-mlL}}{(mlL)^{(D+2)/2}}\sum_{j=1}^\infty j \sinh(j \beta \tilde{\mu}) \nonumber\\
& \times \left \{1+2\sideset {}{'}\sum_{k=1}^{[q/2]}\cos \left( 2\pi k\alpha _{0}\right) -\frac{q}{\pi}\int_{0}^{\infty }dy \ \frac{h(q,\alpha_{0},y)}{\cosh (qy)-\cos (q\pi )} \right \} \ .
  \label{large-L-comp-charge}
\end{align}%
As can be seen, the above expression goes exponentially to zero at this limit.

The total charge density at thermal equilibrium with temperature $T$ reads
\begin{align}
\left\langle j^{0}\right\rangle &=\left\langle j^{0}\right\rangle_M+\frac{8e \beta m^{D+1}}{(2\pi)^{(D+1)/2}}\sideset {}{'}\sum_{l=0}^\infty \cos(2\pi l \tilde{\eta}) \sum_{j=1}^\infty j \sinh(j \beta \tilde{\mu}) \nonumber\\
& \times \left \{ 2\sideset {}{'}\sum_{k=1}^{[q/2]}\cos \left( 2\pi k\alpha _{0}\right) \right.  f_{(D+1)/2}\left(2 m \sigma_{j,l,k}(\beta,q)\right) \nonumber\\
&-\frac{q}{\pi}\int_{0}^{\infty }dy \ \frac{h(q,\alpha_{0},y)}{\cosh (qy)-\cos (q\pi )}\left. f_{(D+1)/2}\left(2 m \delta_{j,l}(\beta,y)\right)\phantom{\sum_{k=1}^{[q/2]}}\hspace{-0.7cm}\right \} \ ,
  \label{total-charge-density}
\end{align}%
where the Minkowskian part ($q=1$ and $\alpha_0=0$) is as follows
\begin{align}
\left\langle j^{0}\right\rangle_M =\frac{8e \beta m^{D+1}}{(2\pi)^{(D+1)/2}}\sideset {}{'}\sum_{l=0}^\infty \cos(2\pi l \tilde{\eta}) \sum_{j=1}^\infty j \sinh(j \beta \tilde{\mu}) f_{(D+1)/2}(m \rho_{j,l}(\beta)) \ ,
  \label{Minkowski-total-charge-density}
\end{align}%
and the prime on the summation over $l$ means that we should take the term $l=0$ with a factor of $1/2$. This term corresponds to the charge density when there is no compactification. Note that, the Minkowskian contribution is independent of the distance from the string and therefore is homogeneous in space.
 
In the massless case, the chemical potential $\tilde{\mu}$ is also zero because of the condition $|\tilde{\mu}|\leqslant m$ and therefore the charge density vanishes.

In the absence of the conical defect, $q=1$, we have
\begin{align}
\left\langle j^{0}\right\rangle &=\frac{8e \beta m^{D+1}}{(2\pi)^{(D+1)/2}}\sideset {}{'}\sum_{l=0}^\infty \cos(2\pi l \tilde{\eta}) \sum_{j=1}^\infty j \sinh(j \beta \tilde{\mu}) \left \{ \phantom{\sum_{k=1}^{[q/2]}}\hspace{-0.7cm} f_{(D+1)/2}(m \rho_{j,l}(\beta)) \right.  \nonumber\\
&-\frac{1}{2\pi}\int_{0}^{\infty }dy \ \frac{h(1,\alpha_{0},y)}{\cosh^2 (y/2)}\left. \phantom{\sum_{k=1}^{[q/2]}}\hspace{-0.7cm} f_{(D+1)/2}\left(2 m \delta_{j,l}(\beta,y)\right)\right \} \ .
  \label{no-con}
\end{align}%
with 
$h(1,\alpha_{0},y)=\sin{(|\alpha_0| \pi)} \left[\cosh{\left(|\alpha_0| y\right)}+\cosh{\left(\left(1-|\alpha_0|\right) y\right)}\right]$.

Now, let us consider the behavior of the charge density in different asymptotic regions of the
parameters. First we investigate the region near the string. In contrast with the fermionic case, which may present a divergent result \cite{JPA}, the charge density is finite on the string. This can be easily obtained by analyzing the integral of the last term in (\ref{total-charge-density}) taking $r =0$. The result is
\begin{align}
\left\langle j^{0}\right\rangle &= \frac{8 e \beta m^{D+1}}{(2\pi)^{(D+1)/2} }  \sideset {}{'}\sum_{l=0}^\infty \cos(2\pi l \tilde{\eta}) \sum_{j=1}^\infty j \sinh(j \beta \tilde{\mu})  f_{(D+1)/2}(m \rho_{j,l}(\beta))\nonumber\\
& \times \left \{ 1+2\sideset{}{'}\sum_{k=1}^{[q/2]}\cos \left( 2\pi k\alpha _{0}\right) -\frac{q}{\pi}\int_{0}^{\infty }dy \ \frac{ h(q,\alpha_{0},y)}{\cosh (qy)-\cos (q\pi )}  \right \} \ .
  \label{near-string-comp-charge}
\end{align}%

At low temperatures, $T\ll m,r^{-1}$, the dominant contribution comes from the term $j=1$ which leads to
\begin{align}
\left\langle j^{0}\right\rangle &\approx\frac{\beta e m^{D+1} e^{\beta \tilde{\mu}}}{(2\pi)^{(D-2)/2} }   \sideset{}{'} \sum_{l=0}^\infty \cos(2\pi l \tilde{\eta}) \left(m \sqrt{\beta^2+l^2 L^2} \right)^{-1/2} e^{-\left(m \sqrt{\beta^2+l^2 L^2} \right)}\nonumber\\
& \times \left \{ 1+2\sideset{}{'}\sum_{k=1}^{[q/2]}\cos \left( 2\pi k\alpha _{0}\right) -\frac{q}{\pi}\int_{0}^{\infty }dy \ \frac{ h(q,\alpha_{0},y)}{\cosh (qy)-\cos (q\pi )}  \right \} \ ,
  \label{zero-temp-comp-densityfin}
\end{align}%
where we have used the asymptotic expansion for large arguments for the MacDonald function as $K_\nu(z)\approx \sqrt{\frac{\pi}{2z}} e^{-z}$. Since $|\tilde{\mu}|\leqslant m$, the above equation tends to zero at the zero temperature limit which is the expected result as we discussed before. When the temperature goes to zero we should retrieve the result obtained in \cite{Eduardo} at zero temperature, which is zero.

At high temperatures or large distances from the string, the main contribution to the charge density in (\ref{total-charge-density}) comes from large $j$ and consequently this representation is not convenient for these limiting cases. In order to find a more convenient
representation in this limit, first we make
the replacement%
\begin{equation}
\beta j \sinh (j\beta \tilde{\mu} )=\partial _{\tilde{\mu} }\cos [i j\beta \tilde{\mu} )],  \label{Repl1}
\end{equation}%
in the part induced by the cosmic string and magnetic flux and then use the
relation \cite{Bell09}
\begin{eqnarray}
&&\sum_{j=-\infty }^{+\infty }\cos (jb)f_{\nu }(m\sqrt{\beta ^{2}j^{2}+a^{2}}%
)=\frac{(2\pi )^{1/2}}{\beta m^{2\nu }}  \notag \\
&&\quad \times \sum_{j=-\infty }^{+\infty }\left[ (2\pi j+b)^{2}/\beta
^{2}+m^{2}\right] ^{\nu -1/2}f_{\nu -1/2}(a\sqrt{(2\pi j+b)^{2}/\beta
^{2}+m^{2}}),  \label{SumHighT}
\end{eqnarray}%
with $b=i\beta \tilde{\mu} $ and $\nu=(D+1)/2$. Using the relation $\left[ x^{2\nu
}f_{\nu }(x)\right] ^{\prime }=-x^{2\nu -1}f_{\nu -1}(x)$, we obtain
\begin{align}
\left\langle j^{0}\right\rangle &=\left\langle j^{0}\right\rangle _{M}-\frac{4 e}{(2\pi)^{D/2} \beta} \sideset{}{'}\sum_{l=0}^\infty \cos(2\pi l \tilde{\eta}) \sum_{j=-\infty}^\infty  (-\tilde{\mu}+2i\pi j/\beta) b_j^{D-2}(\beta) \nonumber\\
& \times \left \{2\sideset{}{'}\sum_{k=1}^{[q/2]}\cos \left( 2\pi k\alpha _{0}\right)\right. f_{D/2-1}\left(\sqrt{l^2 L^2+4r^2 \sin{(\pi k/q)}} b_j(\beta)\right)\nonumber\\
&-\frac{q}{\pi}\int_{0}^{\infty }dy \ \frac{ h(q,\alpha_{0},y)}{\cosh (qy)-\cos (q\pi )} \left.  f_{D/2-1}\left(\sqrt{l^2 L^2+4r^2 \cosh^2 (y/2)} b_j(\beta)\right)\phantom{\sum_{k=1}^{[q/2]}}\hspace{-0.65cm} \right \} \ ,
  \label{high-temp-densityfin}
\end{align}%
where $b_j(\beta)=\sqrt{(2\pi j+i \beta \tilde{\mu})^{2}/\beta^2+m^{2}}$ and the Minkowskinan contribution is given in (\ref{Minkowski-total-charge-density}).
In both limiting cases, high temperatures or large distances from the string, the dominant contribution comes from the terms with $j=-1$ and $j=1$.

\begin{figure}[tbph]
\begin{center}
\begin{tabular}{cc}
\epsfig{figure=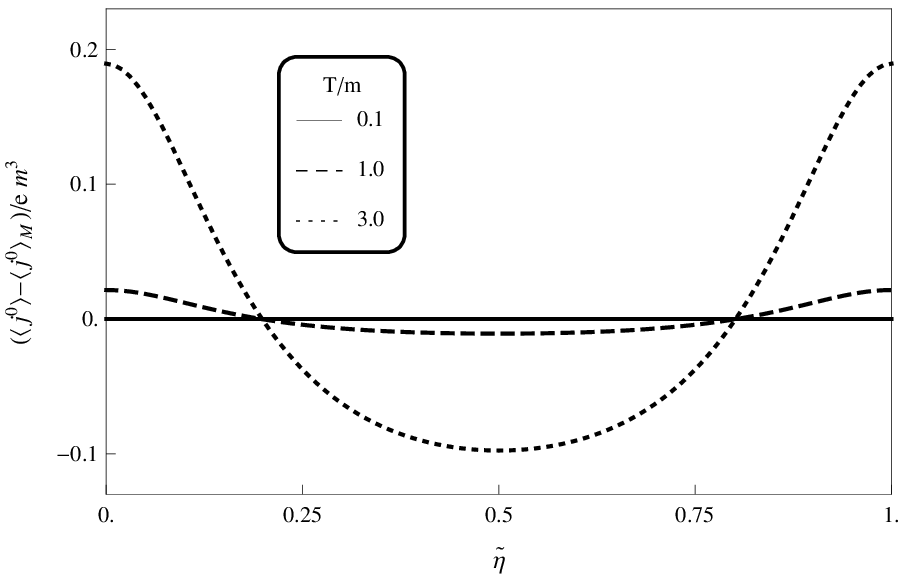,width=0.42\textwidth} & \quad %
\epsfig{figure=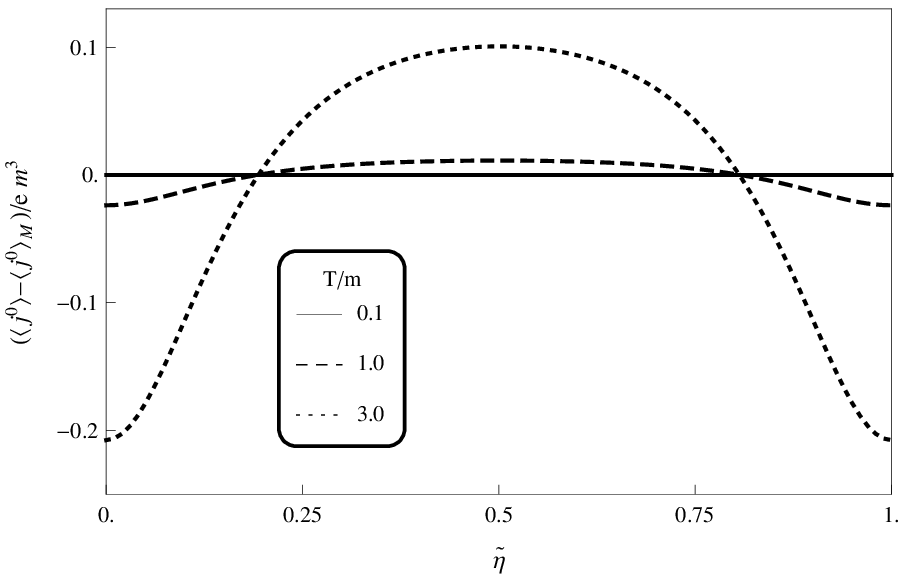,width=0.42\textwidth}%
\end{tabular}%
\end{center}
\caption{The charge density induced by the string and magnetic flux as a function of the parameter $\tilde {\eta}$ for two different values $\alpha _{0}=0$ (left panel) and $\alpha _{0}=0.25$ (right panel). The graphs are plotted for $T/m=0.1,1,3$, $mr=0.5 $, $q=1.5$, $mL=1.0$, $D=3$ and $\protect\mu /m=0.5$. }
\label{fig1Charge}
\end{figure}

The main focus of this paper is to investigate the thermal effect on the charge and current densities. In Figure (\ref{fig1Charge}) we plot the charge density, ignoring the Minkowskian part, as a function of the parameter $\tilde {\eta}$ for three different values of temperature $T/m=0.1, 1$ and $3$. The left panel corresponds to $\alpha _{0}=0$ (in the absence of the magnetic flux) and the right one to $\alpha _{0}=0.25$. As can be seen, the absolute value of the charge density increases with the temperature. Therefore, the temperature intensify the induced charge density. If the contribution of particles (antiparticles) is greater than antiparticles (particles), it will be even greater if one increases the temperature. Comparing the left and right graphs show that depending on the value of $\alpha _{0}$ which is a measure for the magnetic flux, the relative contribution of the particle and antiparticles may reverse. Also, these graphs confirm that the charge density is an even function of $\tilde {\eta}$.

\begin{figure}[tbph]
\begin{center}
\begin{tabular}{cc}
\epsfig{figure=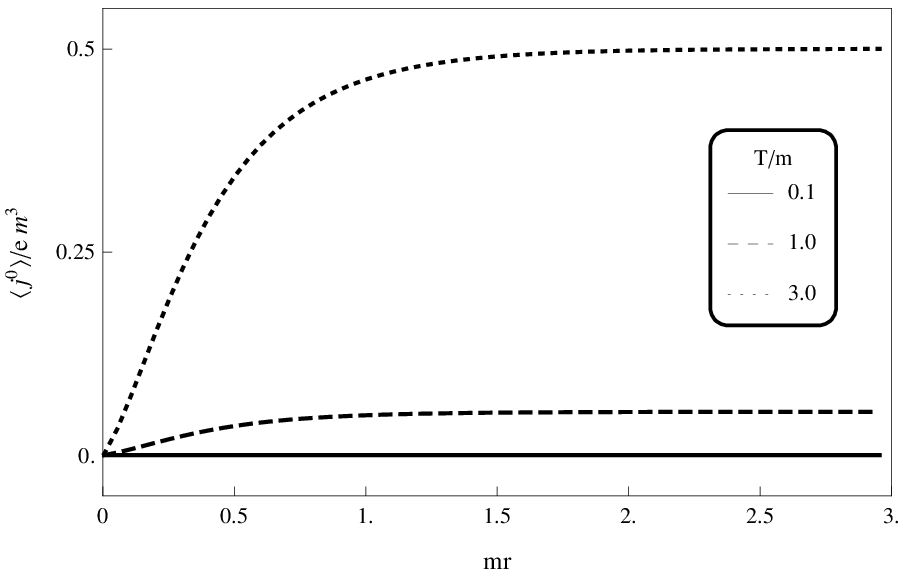,width=0.42\textwidth} & \quad %
\epsfig{figure=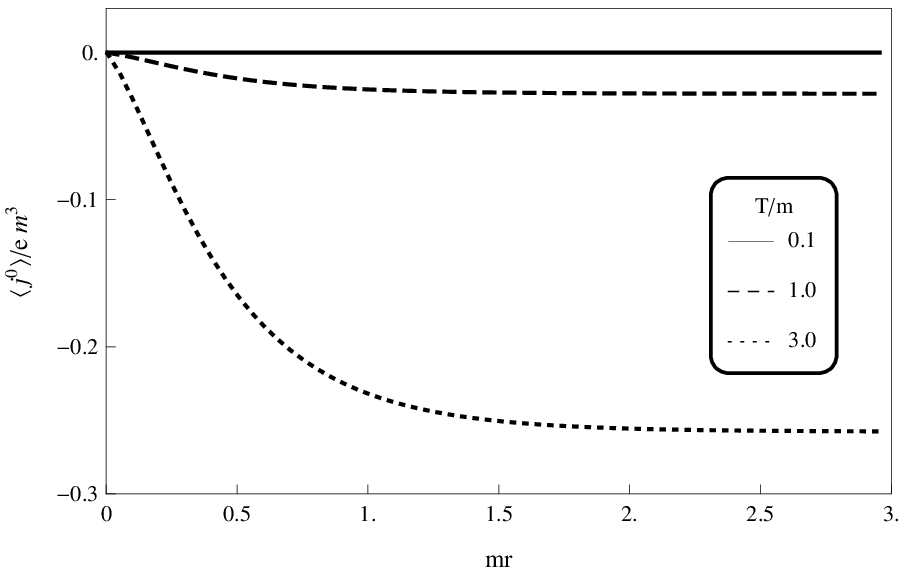,width=0.42\textwidth}%
\end{tabular}%
\end{center}
\caption{The total charge density as a function of the parameter $mr$ for two different values $\protect\tilde{\eta}=0.1$ (left panel) and $\protect\tilde{\eta}=0.7$ (right panel). The graphs are plotted for $T/m=0.1,1,3$, $mr=0.5 $, $q=2.5$, $\protect\alpha _{0}=0.25$, $mL=1.0$, $D=3$ and $\protect\mu /m=0.5$. }
\label{fig3Charge}
\end{figure}

Figure (\ref{fig3Charge}) shows the total charge density as a function of the distance from the string for three different values of temperature $T/m=0.1, 1$ and $3$. The left panel corresponds to $\tilde {\eta}=0.1$ and the right one to $\tilde {\eta}=0.7$. Also here we can see that increasing the temperature from zero, creates nonzero charge density which originates from nonzero chemical potential and as much as the temperature is higher, the effect on the charge density is higher too. At large distances from the string, the effect of the string and the magnetic flux running through it is negligible and the charge density tends to the Minkowskian contribution at temperature T which is constant in space. This can be seen in all curves in this figure which converged to specific values depending on the temperature. Comparing the left and right graphs, it is clear that different values of $\tilde {\eta}$ can reverse the importance of the particle and antiparticle contributions in the charge density. This behavior depends on the value of $\tilde {\eta}$ being less or greater than 0.5. 

\section{Azimuthal current density}
\label{Azimuthal current}
It is not difficult to show that the radial current density, $\left\langle j^{1}\right\rangle$, and the ones in the extra dimensions, $\left\langle j^{i}\right\rangle$ for $i\ge4$, vanish. So, azimuthal and axial current densities are the only ones that are nonzero. First we calculate the induced azimuthal current density, considering $\nu=2$ component in (\ref{Cur}). In this case we have
\begin{equation}
\left\langle j_{2}\right\rangle _{T}=-\frac{e q^2}{(2\pi)^{D-2}  L}\sum_{\sigma } \  \frac{\lambda}{E_l} (n+\alpha) J_{q|n+\alpha|}^2(\lambda r) \ \left[\frac{1}{e^{\beta(E_l-\tilde{\mu})}-1}+\frac{1}{e^{\beta(E_l+\tilde{\mu})}-1}\right] \ .
  \label{16}
\end{equation}%
The corresponding analysis for the vacuum induced azimuthal current was given in \cite{Eduardo}. Here, we shall analyze the thermal contribution. Note that, the azimuthal current density is an odd periodic function of the parameter $\alpha$, which means that in the absence of the magnetic flux along the string it vanishes. Also, it is an even function of the chemical potential $\tilde{\mu}$. In the absence of the chemical potential, the contribution of the particles and antiparticles coincide.

By using the series representation (\ref{Expansion}), the equation for the azimuthal current density reduces to
\begin{equation}
\left\langle j_{2}\right\rangle _{T}=-\frac{2e q^2}{(2\pi)^{D-2}  L}\int d\bold{k}_{||} \int^\infty_0 d\lambda \ \sum_{l=-\infty }^{+\infty
}\frac{\lambda}{E_l}\sum_{n=-\infty}^{+\infty}  (n+\alpha)  J_{q|n+\alpha|}^2(\lambda r)
 \sum_{j=1}^\infty e^{-j \beta E_l} \cosh(j \beta \tilde{\mu}) \ .
  \label{17}
\end{equation}%
Using again the Abel-Plana summation formula, it is possible to separate the string and compactification contributions so as in the previous section. Choosing $g(u)=1$ and taking
\begin{equation}
f(u)=\frac{e^{-j \beta \sqrt{\lambda ^{2}+(2\pi u/L)^{2}+\bold{k}_{||}^2+m^{2}}}}{\sqrt{\lambda ^{2}+(2\pi u/L)^{2}+\bold{k}_{||}^2+m^{2}}}\ ,  \label{fu}
\end{equation}%
the azimuthal current density can be written as
\begin{equation}
\left\langle j_{2}\right\rangle _{T} =\left\langle j_{2}\right\rangle _{Ts}+\left\langle j_{2}\right\rangle _{Tc}\ ,
\end{equation}%
where the string induced part is given by
\begin{align}
\left\langle j_{2}\right\rangle _{Ts}&=-\frac{2eq^2}{\pi(2\pi)^{D-2} }\int d\bold{k}_{||}\int^\infty_0 d\lambda \ \lambda \int^\infty_0 dk \sum_{n=-\infty}^{+\infty} (n+\alpha) J_{q|n+\alpha|}^2 (\lambda r) \nonumber\\
&\times \sum_{j=1}^\infty  \frac{e^{-j \beta \sqrt{\lambda ^{2}+k^{2}+\bold{k}_{||}^2+m^{2}}}}{\sqrt{\lambda ^{2}+k^{2}+\bold{k}_{||}^2+m^{2}}} \cosh(j \beta \tilde{\mu}) \ ,
  \label{18}
\end{align}%
and the compactification induced one as
\begin{align}
\left\langle j_{2}\right\rangle _{Tc}&=-\frac{2eq^2}{\pi (2\pi)^{D-2}}\int d\bold{k}_{||}\int^\infty_0 d\lambda \ \lambda \int^\infty_{\sqrt{\lambda ^{2}+\bold{k}_{||}^2+m^{2}}} dk \sum_{n=-\infty}^{+\infty}  (n+\alpha) J_{q|n+\alpha|}^2 (\lambda r) \nonumber\\
&\times \sum_{j=1}^\infty \cosh(j \beta \tilde{\mu}) \frac{\cos\left({j \beta \sqrt{k^{2}-\bold{k}_{||}^2-\lambda ^{2}-m^{2}} }\right)}{\sqrt{k^{2}-\bold{k}_{||}^2-\lambda ^{2}-m^{2}}} \left[\frac{1}{e^{L k+2\pi i \tilde{\eta}}-1}+\frac{1}{e^{L k-2\pi i \tilde{\eta}}-1}\right] \ .
  \label{19}
\end{align}%

Let us first consider the string contribution in the azimuthal current density. Using the integral representation (\ref{Rel4}) and then integrating over $\lambda$ and $\bold{k}_{||}$ in the same way as in (\ref{Int-reg}) and (\ref{extrad}), we obtain
\begin{align}
\left\langle j_{2}\right\rangle _{Ts}=-\frac{8eq^2}{(4\pi)^{(D+1)/2} } \ \int^\infty_0 ds\  s^{-D} G(q,\alpha_0,r^2/2 s^2)  \sum_{j=1}^\infty  e^{-m^2 s^2-(j^2 \beta^2/4+r^2/2)s^{-2}} \cosh(j \beta \tilde{\mu}) \ ,
  \label{20}
\end{align}%
where we have introduced the notation \cite{Eduardo}
 \begin{align}%
 G(q,\alpha_0,z)& = \sum_{n=-\infty}^{+\infty}\ (n+\alpha) I_{q|n+\alpha|}(z) \nonumber\\
&=\frac{z}{q \pi }%
\int_{0}^{\infty }dy \ \frac{\sinh \left( y\right) e^{-z\cosh {(y)}}g(q,\alpha
_{0},y)}{\cosh (qy)-\cos (q\pi )}   \notag \\
& +\frac{2 z}{q^2}\sideset {}{'}\sum_{k=1}^{[\frac{q}{2}]}\sin{(2\pi k/q)}\sin \left( 2\pi k\alpha _{0}\right)
e^{z\cos (2\pi k/q)} \ ,  \label{roReg}
 \end{align}%
with
\begin{equation}
g(q,\alpha_{0},x)=\sin \left (q \pi \alpha_0\right) \sinh \left[\left(1-|\alpha_0|\right) q y\right]-\sinh \left (q y \alpha_0\right) \sin \left[\left(1-|\alpha_0|\right) q \pi \right] \  .  \label{g}
\end{equation}%
Substituting the above result into (\ref{20}), we can integrate over $s$ and obtain the following result
\begin{align}
\left\langle j^{2}\right\rangle _{Ts}&=\frac{8 e m^{D+1}}{(2\pi)^{(D+1)/2}} \sum_{j=1}^\infty \cosh(j \beta \tilde{\mu}) \left[\ \sideset {}{'}\sum_{k=1}^{[q/2]}\sin{(2\pi k/q)}\sin \left( 2\pi k\alpha _{0}\right)\right. f_{(D+1)/2}\left(2 m c_{j,k}(\beta,q)\right)\nonumber\\
&\left.+\frac{q}{2\pi}\int_{0}^{\infty }dy \ \frac{\sinh \left( y\right) g(q,\alpha_{0},y)}{\cosh (qy)-\cos (q\pi )}f_{(D+1)/2}\left(2 m s_j(\beta,y)\right)\phantom{\phantom{\sum_{k=1}^{[q/2]}}}\hspace{-0.6cm}\right] \ ,
  \label{string-azimuth}
\end{align}%
where $c_{j,k}(\beta,q)$ and $s_j(\beta,y)$ are given in (\ref{fnu}).

Now, we turn to the part induced by the compactification of the string along its axis. To calculate this part we start with the change of variable as
$p=\sqrt{k^2-\bold{k}_{||}^2-\lambda^2-m^2}$ in Eq. (\ref{19}) which gives
 \begin{align}
\left\langle j^{2}\right\rangle _{Tc}&=-\frac{4eq^2}{\pi (2 \pi)^{D-2}}\sum_{l=1}^\infty \cos(2\pi l \tilde{\eta})\int d\bold{k}_{||}\int^\infty_0 d\lambda \  \lambda \int^\infty_0 dp \ \frac{e^{-l L \sqrt{p^2+\bold{k}_{||}^2+\lambda^2+m^2}}}{\sqrt{p^2+\bold{k}_{||}^2+\lambda^2+m^2}}\nonumber\\
& \times \sum_{n=-\infty}^{+\infty} (n+\alpha) J_{q|n+\alpha|}^2 (\lambda r) \sum_{j=1}^\infty \cosh(j \beta \tilde{\mu}) \cos\left({j \beta p }\right) \ .
  \label{22}
\end{align}%
Using the integral representation (\ref{Rel4}) and then performing the integration over $\lambda$ and $\bold{k}_{||}$ with the help of (\ref{Int-reg}) and (\ref{extrad}), besides the integration over $p$ given below
\begin{align}
\int_0^\infty dp \ e^{-p^2 s^2} \cos(j \beta p)=\frac{\sqrt{\pi}}{2s} e^{-j^2 \beta^2/4 s^2} \ ,
\end{align}
we obtain the following general expression 
\begin{align}
\left\langle j^{2}\right\rangle _{Tc}&=\frac{16 e m^{D+1}}{(2\pi)^{(D+1)/2}}\sum_{l=1}^\infty \cos(2\pi l \tilde{\eta}) \sum_{j=1}^\infty \cosh(j \beta \tilde{\mu}) \nonumber\\
& \times \left \{ \sideset {}{'}\sum_{k=1}^{[q/2]}\sin{(2\pi k/q)}\sin \left( 2\pi k\alpha _{0}\right)\right. f_{(D+1)/2}\left(2 m \sigma_{j,l,k}(\beta,q)\right)\nonumber\\
&+\frac{q}{2\pi}\int_{0}^{\infty }dy \ \frac{\sinh \left( y\right) g(q,\alpha_{0},y)}{\cosh (qy)-\cos (q\pi )}\left. f_{(D+1)/2}\left(2 m \delta_{j,l}(\beta,y)\right) \phantom{\sum_{k=1}^{[q/2]}}\hspace{-0.65cm}\right \} \ ,
  \label{comp-azimuth}
\end{align}%
which tends to zero in the limit $L\rightarrow \infty$. 

The total azimuthal current density at temperature $T$ is given by
\begin{align}
\left\langle j^{2}\right\rangle &=\frac{16 e m^{D+1}}{(2\pi)^{(D+1)/2}}\sideset {}{'}\sum_{l=0}^\infty \cos(2\pi l \tilde{\eta}) \sideset{}{'}\sum_{j=0}^\infty \cosh(j \beta \tilde{\mu}) \nonumber\\
& \times \left \{ \sideset {}{'}\sum_{k=1}^{[q/2]}\sin{(2\pi k/q)}\sin \left( 2\pi k\alpha _{0}\right)\right. f_{(D+1)/2}\left(2 m \sigma_{j,l,k}(\beta,q)\right)\nonumber\\
&+\frac{q}{2\pi}\int_{0}^{\infty }dy \ \frac{\sinh \left( y\right) g(q,\alpha_{0},y)}{\cosh (qy)-\cos (q\pi )}\left. f_{(D+1)/2}\left(2 m \delta_{j,l}(\beta,y)\right) \phantom{\sum_{k=1}^{[q/2]}}\hspace{-0.65cm}\right \} \ ,
  \label{total-azimuth}
\end{align}%
where the parameters $\sigma_{j,l,k}(\beta,q)$ and $\delta_{j,l}(\beta,y)$ are the ones introduced in (\ref{rho-sigma-delta}). Also, the prime on the summation over $j$ means the term $j=0$ should be taken by a factor $1/2$. This term corresponds to the charge density at zero temperature. As mentioned in the previous section, the term $l=0$ gives the azimuthal current density in the absence of the compactification, $\left\langle j^{2}\right\rangle_{Ts}$. In the absence of the magnetic flux, the azimuthal current density vanishes which means that the Minkowkian part is zero.

In what follows we shall provide some limiting cases for the total azimuthal current density. In the absence of the conical defect, $q=1$, the azimuthal current density reduces to
\begin{align}
\left\langle j^{2}\right\rangle &=\frac{8 e m^{D+1}}{(2\pi)^{(D+3)/2}}\sideset {}{'}\sum_{l=0}^\infty \cos(2\pi l \tilde{\eta}) \sideset{}{'}\sum_{j=0}^\infty \cosh(j \beta \tilde{\mu}) \nonumber\\
&\times \int_{0}^{\infty }dy \ \frac{\sinh \left( y\right) g(1,\alpha_{0},y)}{\cosh^2 (y/2)} f_{(D+1)/2}\left(2 m \delta_{j,l}(\beta,y)\right)  \ ,
  \label{no-con-azimuth}
\end{align}%
with $g(1,\alpha_{0},x)=\sin \left (\pi \alpha_0\right) \sinh \left[\left(1-|\alpha_0|\right) y\right]-\sinh \left ( y \alpha_0\right) \sin \left(|\alpha_0| \pi \right)$. 

For the massless boson \footnote{Massless boson implies $\mu=0.$}, using the MacDonald function for small arguments, we have
\begin{align}
\left\langle j^{2}\right\rangle &\approx \frac{8 e \Gamma{(\frac{D+1}{2})}}{(2\pi)^{(D+1)/2}}\sideset {}{'}\sum_{l=0}^\infty \cos(2\pi l \tilde{\eta}) \sideset{}{'}\sum_{j=0}^\infty  \nonumber\\
& \times \left \{ \sideset {}{'}\sum_{k=1}^{[q/2]}\sin{(2\pi k/q)}\sin \left( 2\pi k\alpha _{0}\right)\right. \left[2\sigma^2_{j,l,k}(\beta,q)\right]^{-(D+1)/2}\nonumber\\
&+\frac{q}{2\pi}\int_{0}^{\infty }dy \ \frac{\sinh \left( y\right) g(q,\alpha_{0},y)}{\cosh (qy)-\cos (q\pi )}\left.\left[2 \delta^2_{j,l}(\beta,y)\right]^{-(D+1)/2}\phantom{\sum_{k=1}^{[q/2]}}\hspace{-0.65cm} \right \} \ .
  \label{massless-compac-azimuth}
\end{align}%
In the limit $r\rightarrow 0$, discarding the zero temperature part which is divergent in this limit (see the discussion in \cite{Eduardo}), the nonzero-temperature azimuthal current density is finite when $|\alpha_0|\geqslant1/q$. Knowing $|\alpha_0| \le 1/2$, the only possibility is $q\ge2$. In this case we can simply take $r=0$ which results in 
\begin{align}
\left\langle j^{2}\right\rangle_T &=\frac{16 e m^{D+1}}{(2\pi)^{(D+1)/2}} \sideset {}{'}\sum_{l=0}^\infty \cos(2\pi l \tilde{\eta}) \sum_{j=1}^\infty \cosh(j \beta \tilde{\mu}) f_{(D+1)/2}\left(m \rho_{j,l}(\beta)\right) \left[\ \sideset {}{'}\sum_{k=1}^{[q/2]}\sin{(2\pi k/q)}\sin \left( 2\pi k\alpha _{0}\right)\right. \nonumber\\
&\left.+\frac{q}{2\pi}\int_{0}^{\infty }dy \ \frac{\sinh \left( y\right) g(q,\alpha_{0},y)}{\cosh (qy)-\cos (q\pi )}\phantom{\sum_{k=1}^{[q/2]}}\hspace{-0.5cm}\right] \ .
  \label{near-string-convergent-azimuth}
\end{align}%
For the case $|\alpha_0|<1/q$, the dominant contribution to the integral over $y$ comes from large values of $y$.  Replacing the integrand by its asymptotic form, one obtains
\begin{align}
\left\langle j^{2}\right\rangle_T &\approx\frac{16 e q m^{D+1}(m r)^{-2(1-|\alpha_0|q)}}{(2\pi)^{(D+3)/2}} \sideset {}{'}\sum_{l=0}^\infty \cos(2\pi l \tilde{\eta}) \sum_{j=1}^\infty \cosh(j \beta \tilde{\mu}) \nonumber\\
& + \left[2^{-|\alpha_0|q} \sin(q \alpha_0 \pi) \Gamma[1-|\alpha_0|q] f_{(D+1)/2+|\alpha_0|q-1}(m\rho_{j,l}(\beta)) \right. \nonumber\\
& \left. -2^{-(1-|\alpha_0|)q}(m r)^{2q(1-2|\alpha_0|)} \ \text{sign}{(\alpha_0)} \sin(q\pi(1-|\alpha_0|)) \Gamma[1-(1-|\alpha_0|)q] \right.  \nonumber\\
&\left.  
\times f_{(D+1)/2+(1-|\alpha_0|)q-1}\left(m\rho_{j,l}(\beta)\right) \phantom{2^{|\alpha_0|}}\hspace{-0.7cm} \right]  \ .
  \label{near-string-divergent-azimuth}
\end{align}%

At low temperatures, $T\ll m,r^{-1}$, the contributions of $j=0$, zero temperature contribution, and $j=1$ are dominant. Using the large argument asymptotic expansion of MacDonald function, (\ref{total-azimuth}) reduces to
\begin{align}
\left\langle j^{2}\right\rangle &\approx \frac{2 e m^{D+1}}{(2\pi)^{(D-2)/2} }  \sideset {}{'}\sum_{l=0}^\infty \cos(2\pi l \tilde{\eta}) \left[ \left(m l L\right)^{-1/2} e^{-m l L} +\left(m \sqrt{\beta^2+l^2 L^2} \right)^{-1/2} e^{\left(-m \sqrt{\beta^2+l^2 L^2}+\beta \tilde{\mu} \right)} \right]\nonumber\\
& \times \left \{ 2\sideset{}{'}\sum_{k=1}^{[q/2]}\sin{(2\pi k/q)}\sin \left( 2\pi k\alpha _{0}\right) +\frac{q}{2\pi}\int_{0}^{\infty }dy \ \frac{\sinh \left( y\right) g(q,\alpha_{0},y)}{\cosh (qy)-\cos (q\pi )}   \right \} \ ,
  \label{zero-temp-comp-azimuth}
\end{align}%
in which since $|\tilde{\mu}|\leqslant m$, the second term in the square bracket goes to zero at the zero temperature limit, as expected, and the total current is dominated by the zero temperature contribution $\left\langle j^{2}\right\rangle_0$.

To investigate the behavior of the total azimuthal current density at high temperatures or large distances from the string, we need again to find a more suitable representation. In order to do that we make the replacement $\cosh{(j \beta \tilde{\mu})}=\cos{(i j \beta \tilde{\mu})}$.
Following the same procedure as in the case of the charge density, we obtain
\begin{align}
\left\langle j^{2}\right\rangle &=\frac{8 e}{(2\pi)^{D/2} \beta} \sideset{}{'}\sum_{l=0}^\infty \cos(2\pi l \tilde{\eta}) \sum_{j=-\infty}^\infty   b_j^{D}(\beta) \nonumber\\
& \times \left \{\sideset{}{'}\sum_{k=1}^{[q/2]}\sin {(2 \pi k/q)}\sin \left( 2\pi k\alpha _{0}\right)\right. f_{D/2}\left(\sqrt{l^2 L^2+4r^2 \sin{(\pi k/q)}} b_j(\beta)\right)\nonumber\\
&+\frac{q}{2\pi}\int_{0}^{\infty }dy \ \frac{\sinh{(y)} g(q,\alpha_{0},y)}{\cosh (qy)-\cos (q\pi )} \left.  f_{D/2}\left(\sqrt{l^2 L^2+4r^2 \cosh^2 (y/2)} b_j(\beta)\right) \phantom{\sum_{k=1}^{[q/2]}}\hspace{-0.7cm} \right \} \ ,
  \label{high-temp-azimuthal}
\end{align}%
with $b_j(\beta)=\sqrt{(2\pi j+i \beta \tilde{\mu})^{2}/\beta^2+m^{2}}$. In the aforementioned limits the contributions of $j=-1$ and $j=1$ are dominant.

\begin{figure}[tbph]
\begin{center}
\epsfig{figure=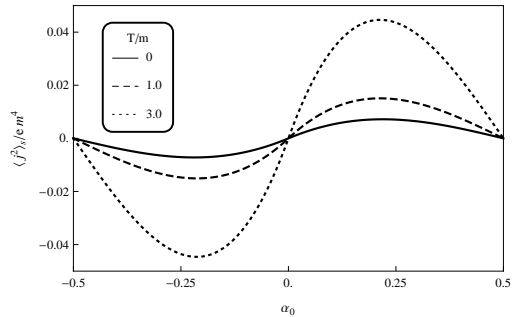,width=0.42\textwidth}
\end{center}
\caption{The azimuthal current density in the absence of compactification as a function of the parameter $\protect\alpha _{0}$. The graphs are plotted for $T/m=0,1,3$, $mr=0.5 $, $q=1.5$, $mL=1.0$, $D=3$ and $\protect\mu /m=0.5$. }
\label{fig1Azimuthal}
\end{figure}

In Figure (\ref{fig1Azimuthal}) we exhibit the behavior of the azimuthal current density in the absence of compactification, i.e. considering only the $l=0$ term in (\ref{total-azimuth}), as a function of the parameter $\alpha _{0}$ for three different values of temperature, $T/m=0,1$ and $3$.
The thermal effect shows the same feature as in the charge density. Increasing the temperature intensifies the current. The value of the current density to be positive or negative is just related to the direction of the current. This graph also confirms that the azimuthal current density is an odd function of $\alpha _{0}$.

\begin{figure}[tbph]
\begin{center}
\begin{tabular}{cc}
\epsfig{figure=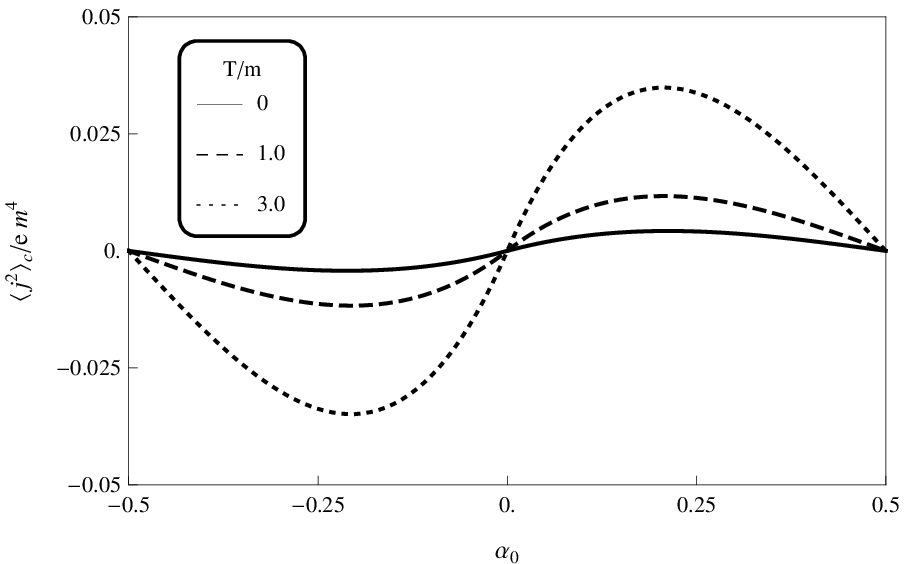,width=0.42\textwidth} & \quad %
\epsfig{figure=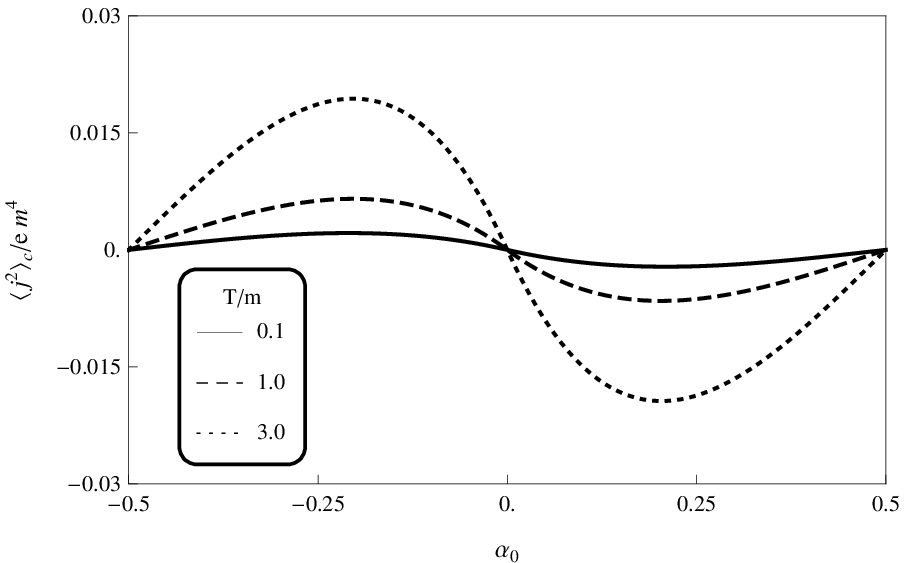,width=0.42\textwidth}%
\end{tabular}%
\end{center}
\caption{The compactification part of the azimuthal current density as a function of the parameter $\protect\alpha _{0}$ for two different values $\protect\tilde{\eta}=0.1$ (left panel) and $\protect\tilde{\eta}=0.7$ (right panel). The graphs are plotted for $T/m=0,1,3$, $mr=0.5 $, $q=1.5$, $mL=1.0$, $D=3$ and $\protect\mu /m=0.5$. }
\label{fig2Azimuthal}
\end{figure}

Figure (\ref{fig2Azimuthal}) shows the compactification part of the azimuthal current density as a function of the parameter $\alpha _{0}$ for three different values of temperature, $T/m=0,1$ and $3$. In the left and right panels, $\tilde{\eta}=0.1$ and $\tilde{\eta}=0.7$ are considered, respectively. As before, increasing the temperature leads to a greater current density. Also, depending on the value of $\tilde{\eta}$ being smaller or greater than 0.5, the direction of the current is inverted. Also, we observe that the intensity of the current increases with $T$.

\begin{figure}[tbph]
\begin{center}
\begin{tabular}{cc}
\epsfig{figure=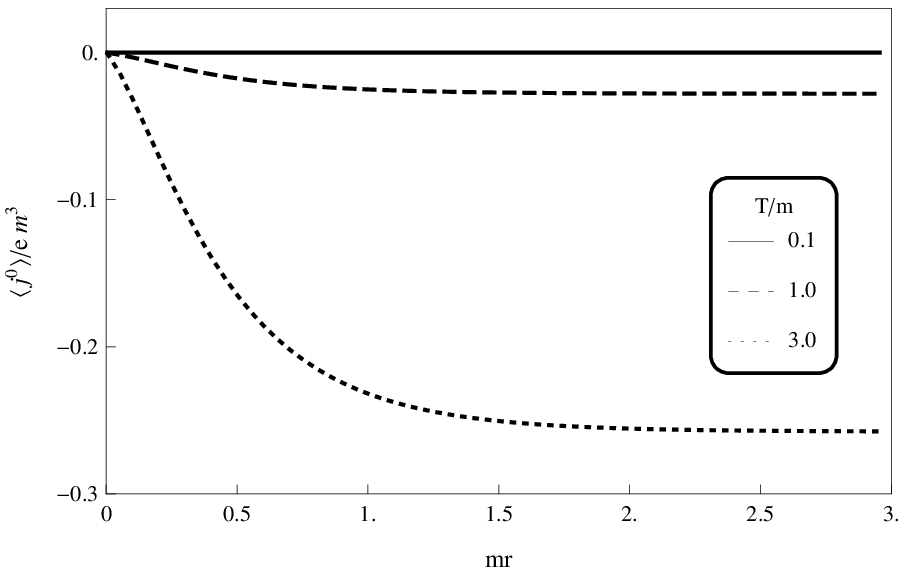,width=0.42\textwidth} & \quad %
\epsfig{figure=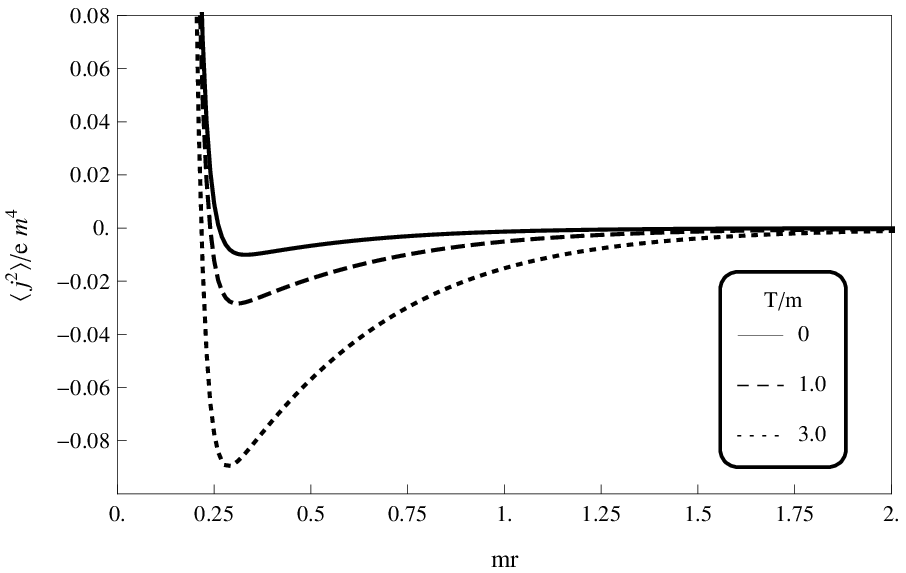,width=0.42\textwidth}%
\end{tabular}%
\end{center}
\caption{The total azimuthal current density as a function of the parameter $mr$ for two different values $\tilde{\eta}=0.0$ (left panel) and $\tilde{\eta}=0.7$ (right panel). The graphs are plotted for $T/m=0,1,3$, $mr=0.5 $, $q=2.5$, $\alpha _{0}=0.25$, $mL=0.1$, $D=3$ and $\mu /m=0.5$. }
\label{fig3Azimuthal}
\end{figure}

In Figure (\ref{fig3Azimuthal}) we plot the total azimuthal current density as a function of the distance from the string for three different values of temperature, $T/m=0,1$ and $3$. Again, increasing the temperature creates greater current density. Also, comparing the left and right graphs which are plotted for two different values of $\tilde{\eta}=0.0$ and $\tilde{\eta}=0.7$ shows different behavior of the current density. As can be seen in the figure, the value of the azimuthal current density goes to the Minkowskian contribution which is zero in this case.

\section{Axial current density}
\label{Axial current}
Now we shall investigate the axial current density which is the $\nu=3$ component in (\ref{Cur}). In this case we have
\begin{equation}
\left\langle j_{3}\right\rangle _{T}=-\frac{eq}{(2 \pi)^{D-2} L}\sum_{\sigma }\ \frac{\lambda \tilde{k_l}}{E_l} J_{q|n+\alpha|}^{2}(\lambda r) \ \left[\frac{1}{e^{\beta(E_l-\tilde{\mu})}-1}+\frac{1}{e^{\beta(E_l+\mu)}-1}\right] \ ,
  \label{31}
\end{equation}%
where $\tilde{k_l}$ is given in (\ref{E+}). With the help of the series expansion (\ref{Expansion}), it results in
\begin{align}
\left\langle j_{3}\right\rangle _{T}&=-\frac{2eq}{(2 \pi)^{D-2} L}\int d\bold{k}_{||}\int^\infty_0 d\lambda \ \sum_{l=-\infty }^{+\infty
}\frac{\lambda \tilde{k_l}}{E_l}\sum_{n=-\infty}^{+\infty} J_{q|n+\alpha|}^{2}(\lambda r) \nonumber\\
&\times \sum_{j=1}^\infty e^{-j \beta E_l} \cosh(j \beta \tilde{\mu}) \ .
  \label{32}
\end{align}%
Again, the summation on the quantum number $l$ can be developed by using the Abel-Plana formula. For this case we take $g(u)=2\pi u/L$ and $f(u)$ as in (\ref{fu}). One can show that $\left\langle j^{3}\right\rangle _{Ts}$ vanishes which is the consequence of the function $g(u)$ being an odd function. So, the only nonzero term is the one induced by the compactification. Therefore, we have
\begin{align}
\left\langle j_{3}\right\rangle _{Tc}&=-\frac{i eq}{\pi (2 \pi)^{D-2} }\int d\bold{k}_{||}\int^\infty_0 d\lambda \  \lambda \int^\infty_{\sqrt{\lambda ^{2}+\bold{k}_{||}^2+m^{2}}} dk\ k \sum_{n=-\infty}^{+\infty}\  J_{q|n+\alpha|}^{2}(\lambda r) \nonumber\\
&\times \sum_{j=1}^\infty  \cosh(j \beta \tilde{\mu}) \frac{\cos\left({j \beta \sqrt{k^{2}-\bold{k}_{||}^2-\lambda ^{2}-m^{2}} }\right)}{\sqrt{k^{2}-\bold{k}_{||}^2-\lambda ^{2}-m^{2}}}\left[\frac{1}{e^{L k+2\pi i \tilde{\eta}}-1}-\frac{1}{e^{L k-2\pi i \tilde{\eta}}-1}\right] \ .
  \label{34}
\end{align}%
After some intermediate steps, following the same procedure as in the previous sections, the total axial current density at temperature $T$ is presented in the form
\begin{align}
\left\langle j^{3}\right\rangle&=\frac{8 e L m^{D+1}}{(2\pi)^{(D+1)/2} } \sum_{l=1}^\infty l \sin(2\pi l \tilde{\eta}) \sideset {}{'} \sum_{j=0}^\infty  \cosh(j \beta \tilde{\mu})  \nonumber\\
& \times \left \{ f_{(D+1)/2}(m \rho_{j,l}(\beta))+2\sideset {}{'}\sum_{k=1}^{[q/2]} \cos \left( 2\pi k\alpha _{0}\right) \right.  f_{(D+1)/2}\left(2 m \sigma_{j,l,k}(\beta,q)\right)\nonumber\\
&-\frac{q}{\pi}\int_{0}^{\infty }dy \ \frac{h(q,\alpha_{0},y)}{\cosh(qy)-\cos (q\pi )} \left. f_{(D+1)/2}\left(2 m \delta_{j,l}(\beta,y)\right) \phantom{\sum_{k=1}^{[q/2]}}\hspace{-0.7cm} \right \} \ ,
  \label{comp-axial}
\end{align}%
where $h(q,\alpha_{0},y)$ is the one introduced in (\ref{h}). Because this current density is induced due to the compactification, in the limit $L\rightarrow \infty$ it vanishes.
Also we can see that this current density is an even function of the chemical potential $\tilde{\mu}$. In the absence of the chemical potential, the particles and antiparticles have the same contributions. 
The equation for the thermal axial current density reduces to the following expression in the absence of conical defect, i.e. $q=1$
\begin{align}
\left\langle j^{3}\right\rangle&=\frac{8 e L m^{D+1}}{(2\pi)^{(D+1)/2} } \sum_{l=1}^\infty l \sin(2\pi l \tilde{\eta}) \sideset{}{'}\sum_{j=0}^\infty  \cosh(j \beta \tilde{\mu})  \nonumber\\
& \times \left \{\phantom{\sum_{k=1}^{[q/2]}}\hspace{-0.7cm} f_{(D+1)/2}(m \rho_{j,l}(\beta)) \right. -\frac{1}{2\pi}\int_{0}^{\infty }dy \ \frac{h(1,\alpha_{0},y)}{\cosh^2 (y/2)} \left. f_{(D+1)/2}\left(2 m \delta_{j,l}(\beta,y)\right) \phantom{\sum_{k=1}^{[q/2]}}\hspace{-0.7cm}\right \} \ .
  \label{no-string-axial}
\end{align}%

Now, we shall analyze the axial current in some specific limits. Let us start with the massless field. So, taking $m=0$, the expression (\ref{comp-axial}) reduces to
\begin{align}
\left\langle j^{3}\right\rangle&= \frac{4 e L \ \Gamma{(\frac{D+1}{2}})}{(2\pi)^{(D+1)/2} }\sum_{l=1}^\infty l \sin(2\pi l \tilde{\eta}) \sideset{}{'}\sum_{j=0}^\infty  \left \{\phantom{\sum_{k=1}^{[q/2]}}\hspace{-0.7cm}\left(\rho^2_{j,l}(\beta)/2\right)^{-(D+1)/2} + \right. \nonumber\\
&\left. 2\sideset {}{'}\sum_{k=1}^{[q/2]} \cos \left( 2\pi k\alpha _{0}\right) \right.  \left[2\sigma^2_{j,l,k}(\beta,q)\right]^{-(D+1)/2}-\frac{q}{\pi}\int_{0}^{\infty }dy \ \frac{h(q,\alpha_{0},y)}{\cosh(qy)-\cos (q\pi )} \left. \left[2\delta^2_{j,l}(\beta,y)\right]^{-(D+1)/2} \phantom{\sum_{k=1}^{[q/2]}}\hspace{-0.7cm}\right \} \ .
  \label{massless-comp-axial}
\end{align}%
by analyzing the integral of the last term in (\ref{comp-axial}) we can see that the axial current density is finite on the string core. The corresponding expression can be obtained by taking $r =0$. The result is
\begin{align}
\left\langle j^{3}\right\rangle&=\frac{8 e L m^{D+1}}{(2\pi)^{(D+1)/2} } \sum_{l=1}^\infty l \sin(2\pi l \tilde{\eta}) \sideset {}{'} \sum_{j=0}^\infty  \cosh(j \beta \tilde{\mu}) f_{(D+1)/2}(m \rho_{j,l}(\beta)) \nonumber\\
& \times \left \{ 1+2\sideset {}{'}\sum_{k=1}^{[q/2]} \cos \left( 2\pi k\alpha _{0}\right) -\frac{q}{\pi}\int_{0}^{\infty }dy \ \frac{h(q,\alpha_{0},y)}{\cosh(qy)-\cos (q\pi )}  \right \} \ .
  \label{near-string-comp-axial}
\end{align}%
Another interesting analysis is related to the behavior of the axial current in the low and high temperature limits.
At low temperatures, $T\ll m,r^{-1}$, the induced axial current density reduces to 
\begin{align}
\left\langle j^{3}\right\rangle&\approx\frac{ L e m^{D+1}}{(2\pi)^{(D-2)/2} }   \sum_{l=1}^\infty  l \sin(2\pi l \tilde{\eta}) \left[\left(m l L \right)^{-1/2} e^{-m l L}+\left(m \sqrt{\beta^2+l^2 L^2} \right)^{-1/2} e^{\left(\beta \tilde{\mu}-m \sqrt{\beta^2+l^2 L^2} \right)} \right]\nonumber\\
& \times \left \{  1+2\sideset{}{'}\sum_{k=1}^{[q/2]}\cos \left( 2\pi k\alpha _{0}\right) -\frac{q}{\pi}\int_{0}^{\infty }dy \ \frac{ h(q,\alpha_{0},y)}{\cosh (qy)-\cos (q\pi )}  \right \} \ ,
  \label{zero-temp-comp-axial}
\end{align}%
where the second term in the square bracket tends to zero in the limit $T\rightarrow 0$.

\begin{figure}[tbph]
\begin{center}
\begin{tabular}{cc}
\epsfig{figure=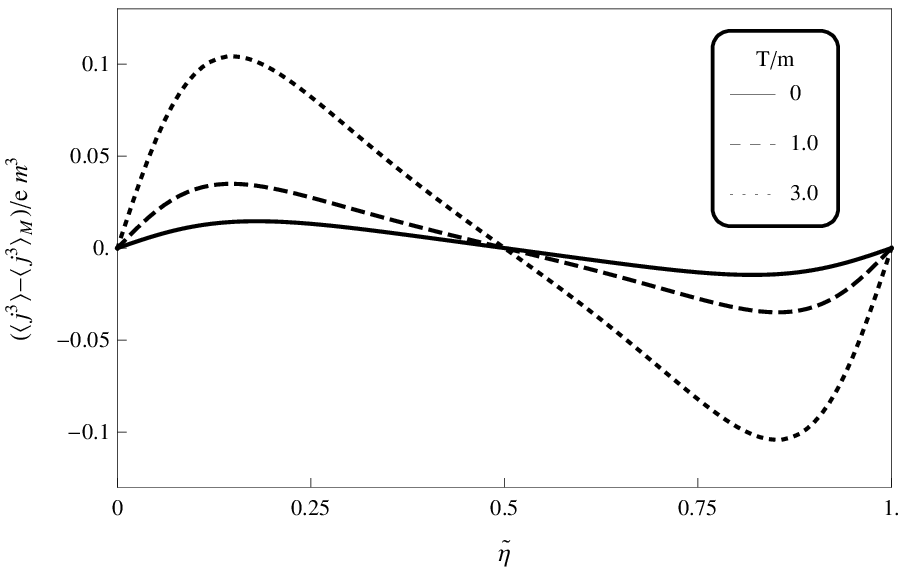,width=0.42\textwidth} & \quad %
\epsfig{figure=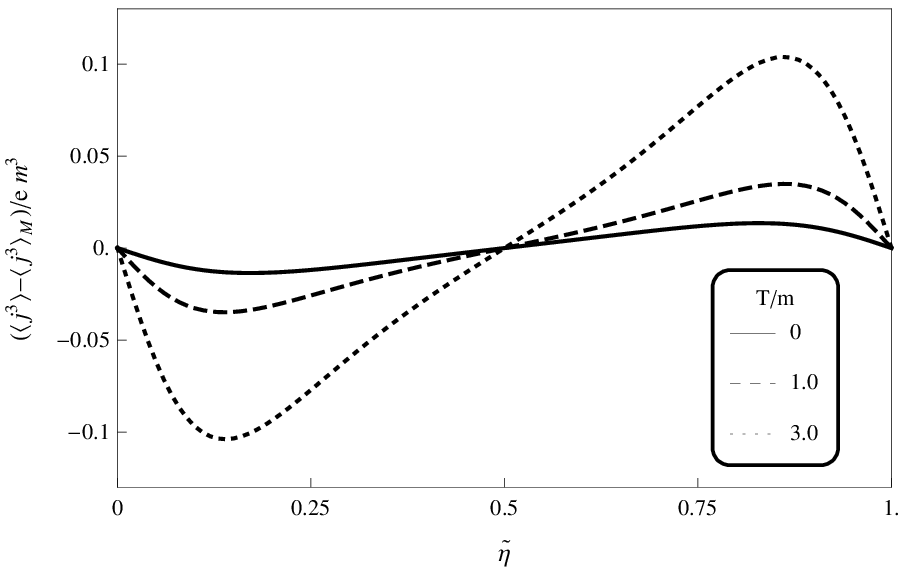,width=0.42\textwidth}%
\end{tabular}%
\end{center}
\caption{The axial current density induced by the string and magnetic flux as a function of the parameter $\tilde {\eta}$ for two different values $\protect\alpha _{0}=0$ (left panel) and $\protect\alpha _{0}=0.25$ (right panel). The graphs on
are plotted for $T/m=0,1,3$, $mr=0.5 $, $q=1.5$, $mL=1.0$, $D=3$ and $\protect\mu /m=0.5$. }
\label{fig1Axial}
\end{figure}

Now, let us investigate the behavior of the axial current density at high temperatures or large distances from the string. As before, we need again to find a better representation. In order to do that we make the replacement $\cosh{(j \beta \tilde{\mu})}=\cos{(i j \beta \tilde{\mu})}$ as in the case of the azimuthal current density. Following the same procedure, we obtain
\begin{align}
\left\langle j^{3}\right\rangle &=\left\langle j^{3}\right\rangle_M + \frac{4 e L}{(2\pi)^{D/2} \beta}\sum_{l=1}^\infty l \sin(2\pi l \tilde{\eta}) \sum_{j=-\infty}^\infty   b_j^{D}(\beta) \nonumber\\
& \times \left \{2\sideset{}{'}\sum_{k=1}^{[q/2]}\sin {(2 \pi k/q)}\sin \left( 2\pi k\alpha _{0}\right)\right. f_{D/2}\left(\sqrt{l^2 L^2+4r^2 \sin{(\pi k/q)}} b_j(\beta)\right)\nonumber\\
&-\frac{q}{\pi}\int_{0}^{\infty }dy \ \frac{h(q,\alpha_{0},y)}{\cosh (qy)-\cos (q\pi )} \left.  f_{D/2}\left(\sqrt{l^2 L^2+4r^2 \cosh^2 (y/2)} b_j(\beta)\right) \phantom{\sum_{k=1}^{[q/2]}}\hspace{-0.7cm} \right \} \ ,
  \label{high-temp-axial}
\end{align}%
where the Minkowskian part ($q=1$ and $\alpha_0=0$) is given by
\begin{align}
\left\langle j^{3}\right\rangle_M =\frac{8e L m^{D+1}}{(2\pi)^{(D+1)/2}} \sum_{l=1}^\infty l \sin(2\pi l \tilde{\eta}) \sideset{}{'}\sum_{j=0}^\infty  \cosh(j \beta \tilde{\mu}) f_{(D+1)/2}(m \rho_{j,l}(\beta)) \ ,
  \label{axial-Minkowskian}
\end{align}%
and the second contribution is induced by the string and magnetic flux.
Figure (\ref{fig1Axial}) exhibit the axial current density induced by the string and magnetic flux as a function of the parameter $\tilde {\eta}$ for three different values of temperature, $T/m=0,1$ and $3$. We plot this current density for two different values of $\alpha_0$; $\alpha _{0}=0$ and $\alpha _{0}=0.25$. They show that the behavior of the current is different in the presence and absence of the magnetic flux. Moreover, the absence or presence of the magnetic flux invert the direction of the axial current density. In this case also the higher temperature provides greater current density. Furthermore, these figures confirm that the axial current density is an odd function of $\tilde {\eta}$.

\begin{figure}[tbph]
\begin{center}
\begin{tabular}{cc}
\epsfig{figure=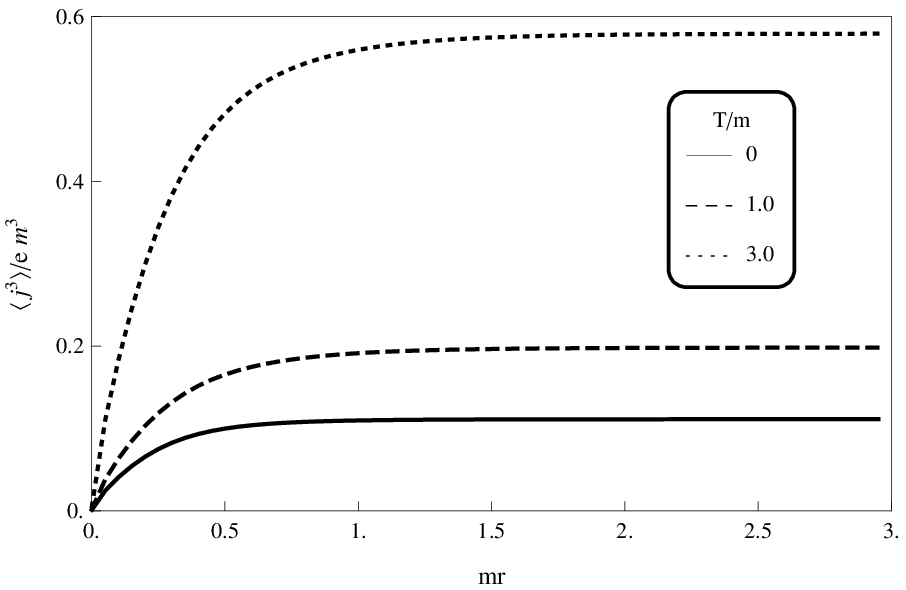,width=0.42\textwidth} & \quad %
\epsfig{figure=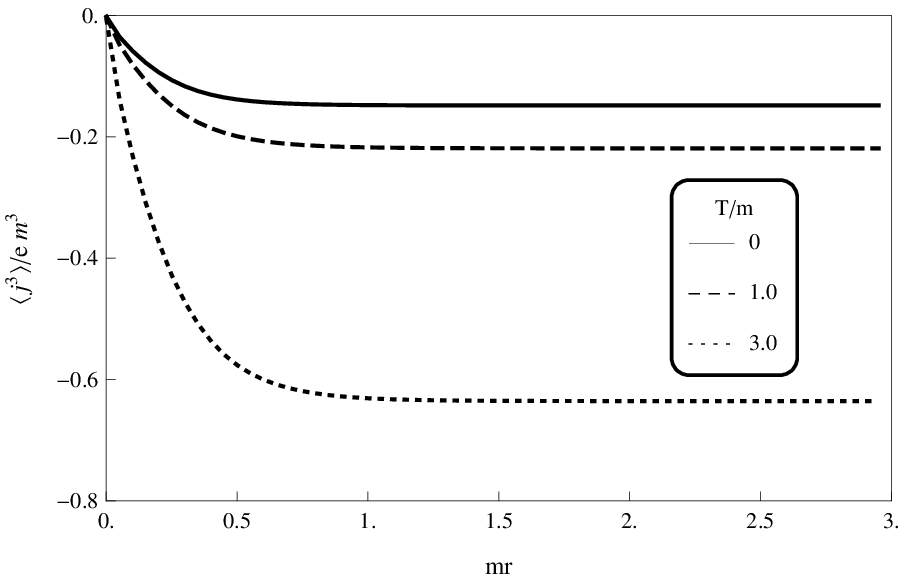,width=0.42\textwidth}%
\end{tabular}%
\end{center}
\caption{The total axial current density as a function of the parameter $mr$ for two different values $\tilde {\eta}=0.1$ (left panel) and $\tilde {\eta}=0.7$ (right panel). The graphs on
are plotted for $T/m=0,1,3$, $\protect\alpha _{0}=0.25$, $q=1.5$, $mL=1.0$, $D=3$ and $\protect\mu /m=0.5$. }
\label{fig2Axial}
\end{figure}

Figure (\ref{fig2Axial}) shows the total axial current density as a function of the distance from the string for three different values of temperature, $T/m=0,1$ and $3$. The current density is plotted for two different values of $\tilde {\eta}=0.1$ and $\tilde {\eta}=0.7$ which are shown in left and right panels, respectively.
As in all other cases, increasing the temperature induces greater current density. All curves in this figure tend to the Minkowskian contribution which is independent of the distance, with the corresponding temperature. Also, depending on the value of the parameter $\tilde {\eta}$ being smaller or greater than 0.5, the direction of the axial current can be inverted.

\section{Conclusion}

\label{conclusion}
In this paper, we have investigated the finite temperature expectation values of the charge and
current densities for a massive bosonic field with nonzero chemical potential in the geometry of a higher dimensional compactified cosmic string with magnetic fluxes, one along the string core and the other enclosed by the compact dimension. In contrast to the fermionic chemical potential which in general can have any value, the bosonic one is restricted by $|\tilde{\mu}|\leqslant \epsilon_0$, $\epsilon_0$ being the minimum of energy. In order to calculate the thermal expectation value of these densities at temperature $T$, we had to calculate the thermal Hadamard function. Working with the grand canonical ensemble and also expanding the field operator over a complete set of normalized positive and negative energy solution, we have decomposed the thermal Hadamard function and consequently the densities to the vacuum expectation values, $\left\langle j^{\nu}\right\rangle_0$, and finite temperature contributions from the particles and antiparticles, $\left\langle j^{\nu}\right\rangle_T$. In the limit $T\rightarrow0$, the latter goes to zero. The charge density is an even periodic function of the magnetic flux with a period equal to the quantum flux and odd function of the chemical potential. Moreover, the azimuthal (axial) current density is an odd (even) periodic function of the magnetic flux with the same period and even function of the chemical potential. We have shown that the bosonic charge and current densities depend only on the fractional part of the ratio of the magnetic flux by the quantum one, $\alpha_0$, which is Aharonov-Bohm-like effect. Also we have shown that although the components of the current densities along the extra dimensions are zero, the charge, azimuthal and axial current densities are affected by the higher dimensions. Thanks to the Abel-Plana formula, we could decompose the densities to the part induced by the string and the one by the compactification. In the limit $L\rightarrow\infty$ the latter vanishes.

For the charge density, the zero temperature expectation value vanishes and finite temperature contribution is given by (\ref{total-charge-density}). When the chemical potential is zero, the contributions from the particles and antiparticles cancel each other and therefore the total charge density vanishes. For the case the bosonic field is massless, the chemical potential is zero and consequently the total charge density vanishes. When the chemical potential is nonzero the particle-antiparticle contributions is not balanced which creates nonzero charge density. In contrast with the fermionic case, which may present a divergent result at the core of the string, the charge density for the bosonic field is finite on the string and has been given in (\ref{near-string-comp-charge}). The behavior of the charge density at low temperature can be obtained directly from (\ref{total-charge-density}) by keeping the terms with $j=1$. The corresponding result is given in (\ref{zero-temp-comp-densityfin}). To investigate the high temperature and large distances from the string an
alternative expression for the charge density, convenient for these limiting cases, is provided in (\ref{high-temp-densityfin}). Increasing the temperature from zero, creates nonzero charge density which originates from nonzero chemical potential. If the contribution of particles (antiparticles) is greater than antiparticles (particles), it will be even greater by increasing the temperature. It means that the temperature intensify the magnitude of the induced charge density. Depending on the value of $\alpha _{0}$, the relative contribution of the particle and antiparticles may reverse. This result is exhibited in Figure (\ref{fig1Charge}).
At large distances from the string, the effect of the string and the magnetic flux running through it is negligible and the charge density tends to the Minkowskian contribution at temperature T which is homogeneous in space. Also, we have shown that different values of $\tilde {\eta}$ can reverse the relevance of the particle and antiparticle contributions in the charge density. This behavior depends on the value of $\tilde {\eta}$ being less or greater than $1/2$. This behavior is displayed in Figure (\ref{fig3Charge}).

The total azimuthal current density which is given by (\ref{total-azimuth}) vanishes in the absence of the magnetic flux. In the limit $r\rightarrow0$, the zero temperature contribution diverges. However, the finite temperature contribution converges when $|\alpha_0|\geqslant1/q$. In this case, we have obtained this contribution of the current density on the string which is given in (\ref{near-string-convergent-azimuth}). For a massless bosonic field the general expression is simplified to (\ref{massless-compac-azimuth}).
 As in the case of the charge density, to investigate the high temperature and large distances from the string a more convenient expression for the azimuthal current density is provided by (\ref{high-temp-azimuthal}).
As before, increasing the temperature leads to a higher current density. Also, it has been explicitly shown in Figure (\ref{fig2Azimuthal}) that depending on the value of $\tilde{\eta}$ being smaller or greater than $1/2$ can invert the direction of the current. We have shown that the value of the azimuthal current density at large distances from the string goes to the Minkowskian contribution which is zero in this case.

The axial current induced by the string is zero and the only nonzero contribution is the one induced by the compactification which is given by (\ref{comp-axial}). For the massless bosonic filed, it reduces to (\ref{massless-comp-axial}). The axial current is finite on the string and can be easily obtained as shown in (\ref{near-string-comp-axial}), taking $r=0$. Again, we have provided a more convenient representation, given by (\ref{high-temp-axial}), to study the high temperature and large distances approximations. As in the other cases, the axial current density increases with the temperature.
We have also shown that the behavior of the system is different in the presence and absence of the magnetic flux. The presence of the magnetic flux can invert the direction of the axial current density, as shown in Figure (\ref{fig1Axial}). 
As before, the axial current density at large distances from the string tends to the Minkowskian contribution with the corresponding temperature which is independent of the distance. Also, depending on the value of the parameter $\tilde {\eta}$ being lower or greater than $1/2$, the direction of the axial current can be inverted.

Finally, we would like to highlight the fact that all the induced charge and current densities present a strong dependence with the temperature. In fact, these quantities are amplified by thermal effects. We can say that this is one of the most important results presented in this paper.

\section*{Acknowledgments}
 
The authors thank Conselho Nacional de Desenvolvimento Cient\'{\i}fico e
Tecnol\'{o}gico (CNPq) for the financial support within the frame of Grant No. 501795/2013-8.

\end{document}